\documentclass{article}

\usepackage{amssymb}

\usepackage{latexsym}
\usepackage[latin1]{inputenc}

\usepackage{graphicx}

\begin{document}

\newfont{\hueca}{msbm10 at 10pt}
\newfont{\griega}{eurm10 at 14pt}
\def\R{\mbox{\hueca R}}
\def\I{\'\i}
\def\h{\langle \cdot, \cdot \rangle}

\newtheorem{defn}{Definition}
\def\thedefn{\thesection.\arabic{defn}}
             
\newtheorem{teo}[defn]{Theorem}
\newtheorem{eje}[defn]{Example}
\newtheorem{lem}[defn]{Lemma}
\newtheorem{obs}[defn]{Remark}
\newtheorem{cor}[defn]{Corollary}
\newtheorem{pro}[defn]{Proposition}

\makeatother

\newcommand{\be}{\begin{equation}}
\newcommand{\ee}{\end{equation}}
\newcommand{\dem}{\noindent {\it Proof. }}

\newcommand{\<}{\langle}
\renewcommand{\>}{\rangle}
\renewcommand{\(}{\left(}
\renewcommand{\)}{\right)}

\newcommand{\om}{\omega}
\newcommand{\Om}{\Omega}
\newcommand{\al}{\alpha}
\newcommand{\bt}{\beta}
\newcommand{\la}{\lambda}
\newcommand{\ga}{\gamma}
\newcommand{\ep}{\epsilon}
\newcommand{\LL}{{\cal L}}

\newcommand{\gal}{G_m(\R )} 
\newcommand{\Z}{{\cal Z}(M)} 
\newcommand{\lei}{{\rm Leib}(M)} 
\newcommand{\leiu}{{\rm Leib}_1(M)} 
\newcommand{\leic}{{\rm Leib}_0(M)} 
\newcommand{\egal}{(M,\Omega, \h, \nabla)} 

\title{\Huge Leibnizian, Galilean and Newtonian structures of spacetime}
\author{Antonio N. Bernal and Miguel S\'anchez\footnote{Part of the results of this article has been announced at the RSME meeting ``Encuentros de Oto\~no de Geometr\'{\i}a y F\'{\i}sica'' Miraflores de la Sierra (Madrid), September 2001.}}

\maketitle
\vspace*{-6mm}
\begin{center}
{\small Dpto. de Geometr\'{\i}a y Topolog\'{\i}a, Universidad de Granada, \\Facultad de Ciencias, Fuentenueva s/n, E--18071 Granada, Spain. \\Email: sanchezm@goliat.ugr.es}

\bigskip

\textbf{Abstract} \\[.3cm]

\end{center}

\begin{quote}

{\small 
The following three geometrical structures on a manifold are studied in detail: 

 Leibnizian: a non-vanishing 1-form $\Omega$ plus a Riemannian metric $\h$ on its annhilator vector bundle. In particular, the possible dimensions of the automorphism group of a Leibnizian $G$-structure are characterized.

 Galilean: Leibnizian structure endowed with an affine connection $\nabla$ (gauge field) which parallelizes $\Omega$ and $\h$. Fixed any vector field of observers $Z$ ($\Omega (Z)\equiv 1$), an explicit Koszul--type formula which reconstruct bijectively all the possible $\nabla$'s from the gravitational  ${\cal G} := \nabla_Z Z$ and vorticity  $\omega:= (1/2)$  rot $Z$ fields (plus eventually the torsion) is provided.

 Newtonian: Galilean structure with $\h$ flat and a field of observers $Z$  which is inertial (its flow preserves the Leibnizian structure and $\omega  \equiv 0$).  Classical concepts in Newtonian theory are revisited and discussed.
}

\end{quote}




\section{Introduction}

It is well-known since Cartan's era \cite{Car} 
that Newtonian theory can be stated in the language of Differential Geometry, and many authors have studied this geometrization in its own and in comparison with (or as a limit of) 
Einstein's General Relativity, see for example, \cite{Di}, 
\cite{DBKP}, \cite{Eh}, \cite{EB}, \cite{Gau}, \cite{KeLe}, \cite{Ku}, \cite{Kub}, 
\cite[Box 12.4]{MTW}, \cite{NS}, \cite{Reb}, \cite{RSB}, \cite{SoMa}, \cite{Tr}. Aim of this article is to carry out a geometrization from a more general viewpoint, which arises from the fundamental considerations on measurements of space and time in \cite{BLS}.

\smallskip

\noindent 
A {\em Leibnizian} structure on a $m$-manifold $M$ is a pair $(\Omega, \h)$ consisting of a 
non-vanishing 1-form $\Omega$ and a (positive definite) Riemannian metric $\h$ 
on its kernel. When  $m=4$, this structure appears naturally as a consequence of our methods of measurement of spacetime; in fact, it is natural to assume the existence of a Leibnizian (or dual {\em anti--Leibnizian}) structure in the degenerate part of a signature-changing metric from Lorentzian to Riemannian \cite{BLS}. When  $\Omega$ is exact, i.e. $\Omega = dT$ for an {\it  absolute time} function $T$,  the intuitive idea ``at each instant of time there exist a Riemannian metric on space'' is geometrized. Fixed the Leibnizian spacetime a {\em Galilean} connection is an affine connection which parallelizes $\Omega$ and $\h$. As a difference with the Levi-Civita connection for a semi-Riemannian (Riemannian, Lorentzian or with any index) manifold, symmetric Galilean connections are not univocally determined by the Leibnizian structure. Moreover, there exist a symmetric Galilean connection if and only if 
$\Omega $ is closed  (i.e., locally, $\Omega = dT$). Galilean connections can be seen 
as gauge fields, which are necessary to preserve the covariance of physical laws under the change of ``Galilean reference frames''.
A Newtonian spacetime will be a Galilean one $(M, \Omega, \h, \nabla)$ where $\nabla$ satisfies certain symmetries.
In the present article we study the mathematical properties of each level
(Leibnizian / Galilean / Newtonian) and the corresponding physical interpretations.

From the purely mathematical viewpoint, some questions arise naturaly:
which are the possible dimensions of the group of automorphisms of a Leibnizian spacetime?,
how many Galilean connections admit a Leibnizian structure?, is there an explicit way to construct them? The answers to such questions are interesting also from the physical viewpoint.
The cornerstone of our approach can be stated as follows (see Lemma \ref{lfk}, Theorem \ref{tfun}, Corollary \ref{cfunbis}): {\em given a Leibnizian spacetime, a field of observers $Z$ and an (unknown) Galilean connection $\nabla$, the gravitational field ${\cal G}$ and the vorticity/Coriolis field $\omega$ measured by $Z$ {\em (plus, eventually, any skew--symmetric tensor Tor representing the torsion, subject to the restriction $\Omega \circ $ Tor $ = d\Omega$)}, permit to reconstruct univocally the connection $\nabla$.} 
Even though partial versions of this result are well-known ({\em ad nauseam} if $\h$ is flat
and $Z$ determines an ``inertial reference frame''), the full result is new, as far as we know.  
In fact, it relies on formula (\ref{fZa}), which plays a similar role to Koszul's formula in semi-Riemannian Geometry, and introduces a type of ``sub--Riemannian'' geometry with interest in  its own. Then,  classical Newtonian concepts are revisited under this viewpoint.

\smallskip

\noindent In the comparison with classical geometrizations of Newtonian theory (see e.gr. \cite{Tr}, \cite[Box 12.4]{MTW}, \cite{Eh}) --where one assumes first that the space is flat and then some sort of assumptions to make inertial references frames  appear--, the advantages of our approach become apparent   not only for its bigger generality but also for  the sake of clarity: the detailed study of the structures at each level Leibnizian/ Galilean/ Newtonian clarifies both the mathematical results and the physical interpretations. It is also worth pointing out that Kunzle and some coworkers \cite{Ku}, \cite{Kub}, \cite{DBKP} have also studied some Leibnizian structures; in fact, 
they call  $(M,\Omega,\h)$ with $\Omega$ closed
 ``Galilei structure" and the corresponding compatible connections  ``Galilei connections''. 
Nevertheless, our constructive procedure  of all Galilean connections and associated physical interpretations go further\footnote{In fact, our study led us to put different names to the structures depending on if $\nabla$ was fixed or not, as in \cite{Eh}. The names Leibnizian, Galilean and Newtonian are suggested by some famous historical facts --Galilean studies on freely moving bodies, controversy between Leibniz and Newton, Newton's discussion of the spinning water-bucket.} (see Remark \ref{ofun}(2)).

\smallskip

\noindent The present article is divided into three parts. In the first one, the properties of pure Leibnizian structures are studied. Leibnizian vector fields and fields of Leibnizian observers (FLO) are introduced, as infinitesimal generators of automorphisms.
In Theorem \ref{tlei}, the possible dimensions of these vector fields are characterized, in agreement to some known properties of classical {\em kinematical group}. 

The second part is devoted to Galilean structures. Apart from the commented results on our Koszul--type formula (\ref{sZa}), we introduce both Galilean vector fields and fields of Galilean observers (i.e. the corresponding Leibnizian fields which preserve infinitesimally the connection $\nabla$), see Table 1. In Section \ref{sfcgc}, coordinate expressions for the connection, geodesics and curvature (for coordinates adapted to general fields of observers as well as more restricted ones: Leibnizian, Galilean or {\em inertial}) are also provided.

Finally, in the third part the Newtonian case is especifically revised, discussing the classical concepts. In fact, our definition of Newtonian spacetime is a Galilean one which admits an inertial field of observers and with $({\rm an}\Omega, \h)$ flat. This definition avoids 
conditions at infinity, which are discussed in relation to the properties of gravitational fields and the uniqueness of Poisson's equation.
Even though from the mathematical viewpoint the results are clearer when non--symmetric connections are also taken into account (see Remark \ref{ofun}(1)), we restrict to symmetric connections for physical concepts or coordinate expressions --in particular along all the third part.

\vspace{1cm}
\begin{quote}
\begin{center}
{\bf Insert Table 1}
\end{center}
\end{quote}

\newpage

\vspace{3mm}
\begin{quote}
\begin{center}
{\bf {\large
I.  LEIBNIZIAN STRUCTURES }}
\end{center}
\end{quote}

\section{Leibnizian spacetimes} \label{seclei}

{\bf {\sc A. Setup.}} A \emph{Leibnizian} spacetime is a triad, $(M,\Omega, \h)$, consisting of a smooth connected manifold $M$,\footnote{As usual, $M$ will be assumed Hausdorff and paracompact;
``smooth'' will mean   
$C^\infty$ (even though $C^2$ is enough).} of any dimension
$m=n+1 \geq 2$, a differential 1-form $\Omega\in\Lambda^1(M)$, nowhere null ($\Omega_p\not=0$, $\forall p\in M$), and a smoth, bilinear, symmetric and positive definite map 
$$\h:\Gamma ({\rm an}\Omega) \times \Gamma({\rm an}\Omega)\longrightarrow C^\infty(M), \quad(V,W) \mapsto \left<V,W\right>,$$  where ${\rm an}\Omega=\{v\in TM\; \mid \;\Omega(v)=0\}$, is the $n$--distribution induced by  $\Omega$, and the symbol $\Gamma$ denotes the corresponding vector fields, so: $\Gamma({\rm an}\Omega) = \{ V \in \Gamma(TM) \mid V_p \in {\rm an}\Omega, \, \forall p \in M \}$. Summing up, the Leibnizian structure on $M$ is the non-vanishing 1-form  $\Omega$ plus the Riemannian vector bundle $({\rm an} (\Omega), \h)$.

\smallskip 

\noindent {\bf  Note.} Let the superscript  $^*$ denote dual space. For any  $p\in M$ there exist a canonical isomorphism between $({\rm an}\Omega_p)^*$ and the quotient vector space $(T_pM)^*/{\rm Span}\Omega_p$. Therefore, the metric $\h_p$ induces a canonical Euclidean product on
$(T_pM)^*/{\rm Span}\Omega_p$, as well as a positive semidefinite metric on 
 $(T_pM)^*$, with radical generated by $\Omega_p$. Thus, a Leibnizian structure is equivalent
to a degenerate semidefinite positive metric of constant rank $n$ in the cotangent bundle $T M^*$, plus a 1-form generating its radical. In \cite{BLS}, an 
{\it anti-Leibnizian} structure on $M$ is defined as  
a degenerate semidefinite positive metric of constant rank $n$ in the tangent bundle $T M$, plus a vector field $Z$ generating its radical. Thus, the study of anti-Leibnizian structures is analogous (dual) to the study of the Leibnizian ones.

\smallskip

\noindent According to \cite{BLS}, Euclidean space $({\rm an}(\Omega_p),\h_p)$ is called the  \emph{absolute space} at $p\in M$, and the linear form $\Omega_p$ is the  \emph{absolute clock} at $p$. A tangent vector $Z_p\in T_pM$ is \emph{timelike}, if $\Omega_p(Z_p)\not=0$ 
(\emph{spacelike}, otherwise). If, additionally, $\Omega_p(Z_p)>0$ (resp. $\Omega_p(Z_p)<0$), $Z_p$ \emph{points out the future} (resp. \emph{the past}) . Any normalized timelike vector $Z_p$ (that is, with  $\Omega_p(Z_p)=1$) is a 
\emph{standard timelike unit} (or \emph{instantaneous observer}) at $p$; any (ordered) orthonormal base of the absolute space at $p$, is a  \emph{set of standard spacelike units} at $p$.

Let us introduce definitions  for the concepts of observer and field of observers (or reference frame) analogous to the Lorentzian ones; compare with \cite[Chapter 2]{SW}. An \emph{observer} is a smooth curve, $\gamma:I\longrightarrow M,$ ($I\subseteq \R$, interval) such that its velocity is always a standard timelike unit,  $\Omega_{\gamma(s)}(\gamma'(s))=1$, $\forall s\in I$. The  parameter of this curve is the \emph{proper time} of the observer $\gamma$. A {\it field of (instantaneous) observers} (FO) is a vector field $Z \in \Gamma (TM)$ with $\Omega(Z) \equiv 1$, that is, integral curves of $Z$ are observers. 
The existence of a FO on any Leibnizian spacetime is straightforward from the paracompactness of $M$.\footnote{Conversely, if we assume the existence of a FO then Lemma \ref{lfk} and Remark \ref{oza} permit us to construct an affine connection on $M$; thus, we could deduce the paracompactness of $M$ by using \cite[vol. II, Addendum 1, p. 8-52]{Sp}.} 
Let $\Z \equiv {\cal Z}(M,\Omega )$ be the set of all the FO's. Clearly, $\Z$ has a structure of affine space with associated vector field $\Gamma ({\rm an}\Omega)$. 
For each FO, $Z\in \Z$, define the field of endomorphisms:

\begin{equation} \label{epe}
P^Z(v)=v-\Omega(v)Z, \;\,\;\forall v\in TM,
\end{equation}
or \emph{spacelike projection along} $Z$. Obviously, the image of $P^Z$ 
is an$(\Omega)$. 

When the absolute clock   $\Omega$ satisfies 
 $\Omega\wedge d\Omega=0$ (i.e. the distribution an$\Omega$ is involutive: $[V,W]\in 
\Gamma({\rm an} \Omega), \, \forall V,W \in \Gamma({\rm an} \Omega))$, we say that $(M,\Omega,\h)$ is {\it 
locally sincronizable}; 
if  $d\Omega=0$ ($\Omega$ is closed), then $(M,\Omega,\h)$ is {\it 
proper time locally syncronizable}. 
In fact, it is well-known that the equality   $\Omega\wedge d\Omega=0$ is equivalent, locally, to   $\Omega=f\,dt$, for some smooth functions $f>0,t$. That is, in the domain of $f$ and $t$, 
hipersuperfaces  $t\equiv $constant are tangent to the absolute space at each point. Thus, in principle, any observer could be
``syncronized'',
that is, it can regard $t$ as a compromise time, obtained by re-scaling its proper time. In the more restrictive case $d\Omega =0$, one has locally $\Omega=dt$. Thus, any observer $\gamma$  
is direcly ``syncronized'', up to a constant $c_{\gamma}$ 
(i.e., $t\circ \gamma(s)= s + c_{\gamma}, \forall s \in I$). Notice that these concepts about local syncronizability are intrinsic to the Leibnizian structure and, then, applicable to each particular observer $\gamma$. This is a clear difference with the Lorenzian case, where the analogous concepts have meaning only for  fields of observers\footnote{If $(M,g)$ is a time-oriented Lorentzian manifold, a FO is a unit future-pointing timelike vector field $Z$. If $Z^{\flat}$ is the metrically associated 1-form,  $Z$ is said locally syncronizable  (resp. proper time locally  sincronizable) if  
$Z^{\flat}\wedge dZ^{\flat} = 0$ (resp. $dZ^{\flat} =0$). It is not difficult to prove that, in the neighborhood of any point, {\it a proper time locally syncronizable vector field can be always constructed}
(compare with \cite[Section 2.3]{SW}).
}.

When  $\Omega$ is exact, that is, $\Omega = dT$ for some (unique up to a constant) $T\in C^\infty (M)$,    $T$  will be called the  function {\em absolute time}.
In this case, any observer $\gamma$ will be assumed to be parametrized with $T$,  
($T\circ \gamma (s) = s, \forall s \in I$). When $M$ is simply connected, 
local proper time synchronizability is equivalent to the existence of such an absolute time function.

\vspace{2mm}

\noindent {\bf {\sc B. Coordinates.}}
Given a Leibnizian spacetime $(M,\Omega,\h)$ and a FO, $Z \in \Z$, 
for each $p\in M$ there exist  charts $(U, y^0,\dots , y^n)$ such that $\partial_{y^0}=Z|_U$. We can wonder if, additionally, these charts may be adapted to the absolute spaces. More precisely:

\begin{defn} \label{defsca}
Let $(M,\Omega,\h)$ a Leibnizian spacetime and  $(U', t,x^1,\dots,x^n)$ a coordinate system in $M$. $(U', t,x^1,\dots,x^n)$ is {\it adapted to the absolute space} if: $$ \Omega(\partial_{x_i})= 0, \quad \quad \forall i\in \{1,\dots,n\}$$ (in particular, hipersuperfaces  $t\equiv$constant are integral manifolds of the distribution ${\rm an}\Omega$).
 
Given a FO, $Z\in \Z$,  $(U', t,x^1,\dots,x^n)$ is {\it adapted to $Z$} if,
on $U'$: 
\[
\partial_t=Z\;\,\; and \;\,\; \Omega = dt.
\]
\end{defn}
If the chart is adapted to the absolute space then
 $\Omega = \Omega(\partial_t) dt$; if it is adapted to $Z$ then it is adapted to the absolute space too. Clearly, if $(U', t,x^1,\dots,x^n)$ is adapted to the absolute space (resp.
a $Z$) then $\Omega \wedge d\Omega = 0$ (resp. $d\Omega =0$) on $U'$. The converse also holds; in fact, the following result yields adapted charts constructively.

\begin{pro} \label{l2.1}
Let $Z$  be a FO on a Leibnizian spacetime  $(M,\Omega,\h)$. Fixed
a chart $(U, y^0,\dots , y^n)$  such that  $\partial_{y^0}=Z|_U$, put 
\be \label{dv}
V_k= P^Z(\partial_{y^k}) \in {\rm an}\Omega, \quad  \forall k\in \{ 1,\dots ,n \}.
\ee
with $P^Z$ in (\ref{epe}). Then:

(i)  $(Z,V_1,\dots,V_n)$ is a local base of vector fields  (\emph{moving frame}) with
$\Omega(V_k)=0$ and:
\begin{equation} \label{corchete}
d\Omega (Z,V_j) = - \Omega([Z,V_j]), \quad \quad d\Omega (V_i,V_j) = - \Omega([V_i,V_j]), \quad \quad \forall i,j\in \{1,\dots n\}.
\end{equation}

(ii) If $\Omega\wedge d\Omega=0$, then, at some neighbourhood
 $U'$ of each $p\in U$, there exist  coordinates
 $(t,x^1,\dots,x^n)$ satisfying on $U'$:
$$\Omega=\Omega(\partial_t)dt, \quad  \partial_{x^k}=V_k, \;\,\;\forall k\in\{1,\dots, n\}.
$$
Thus, such coordinates are adapted to the absolute space.

(iii) If $d\Omega=0$, then, in addition to (ii) one has:
\[
\partial_t=Z,
\]
on $U'$ (i.e. the coordinates are  adapted to $Z$).
\end{pro}

\noindent  \dem  (i) Obvious.

(ii) As the  distribution an$\Omega$ is involutive, $\Omega ([V_i,V_j])=0$ and, from (\ref{dv}), $[V_i,V_j]=0$. Thus, it is enough to apply  classical Frobenius' theorem
(see for example \cite[Ch. 1]{Wa}).

(iii) By using (\ref{corchete}), one checks $[Z,V_j]=0$ and, again, the result follows from Frobenius' theorem. $\square$

From now on, latin indexes $i,j,k$ will vary in $1,\dots,n$. We will simplify the notation  too: $\partial_{x^k} \equiv \partial_k$.

\vspace{2mm}

\noindent {\bf {\sc C. Galilean Group at a point.}} Fixed $p \in M$, 
an (ordered) base $B=(Z_p, e_1, \dots e_n)$ of $T_pM$ is a {\it Galilean base at} $p$ if $\Omega(Z_p) = 1$ and $\{e_1, \dots e_n\}$ is an orthonormal base of an$(\Omega_p$), that is, if  $Z_p$ is a standard timelike unit at $p$ and 
$e_1, \dots e_n$ are standard spacelike units. 

A {\it Galilean transformation at} $p$ is 
a linear map, $A:T_pM\longrightarrow T_pM$, which maps 
some (and thus, any) Galilean base onto a Galilean base. Or, equally,
 $\Omega_p\left(A(X_p)\right)=\Omega_p(X_p)\;\;and\;\left<A(V_p),A(W_p)\right>_p=\left<V_p,W_p\right>_p,\;\forall X_p\in T_pM,\;\forall V_p,W_p\in {\rm an}(\Omega_p)$. The group of all such transformations will be called the
{\it Galilean group } at $p$.

 {\it Matricial Galilean group} $\gal$, $m=n+1$, is the group of the matrixes:
\begin{equation} \label{ggal}
\(
\begin{array}{c|c}
1 & 0 \\ \hline
a   & A
\end{array}
\), \quad \quad \hbox{where} \quad 
a= \left(  
\begin{array}{c}
a^1 \\
\vdots \\
a^n   
\end{array}
\right) \in \R^n
 \quad \hbox{and} \quad A \cdot A^t = I_n 
\end{equation}
($A$ is an orthogonal matrix $n\times n$). 
 
It is straightforward to check that, given a Galilean base $B$ and any other base $B'=(Z'_p, e'_1, \dots e'_n)$ in $T_pM$, the base $B'$ is Galilean if and only if the transition matrix belongs to
$\gal$, that is:
$$
Z'_p = Z_p + \sum_{i=1}^n a^i e_i , \quad \quad e'_j = \sum_{i=1}^n a^i_j e_i, \; \forall j \in \{1, \dots n\} 
$$
where $A=(a^i_j)$ is a  orthogonal matrix. In this case,  $v= \sum_j a^j e_j$ is the {\em velocity of $Z'_p$ measured by} $Z_p$.

\section{Leibnizian vector fields} \label{secalei}

{\sc A. Automorphisms of Leibnizian $G$-Structures.} Let $LM$ be the linear frame bundle of $M$, that is, each element of $LM$ can be seen as a (ordered) base of the tangent space at some point of $M$. The Leibnizian structure  $(\Omega, \h )$ on $M$ determines the fiber bundle of all the Galilean  bases  $GM \subset LM$. As $\gal$ acts freely and transively on each fiber,  $GM$  is a 
$G$--structure with $G= \gal $ (i.e., a principal fiber bundle  with structural group  $\gal$, obtained as a reduction of $LM$). Recall that the set of the orthonormal bases for any 
semi-Riemannian metric (in particular, Riemannian or Lorentzian) is a well--known example of $G$--structure; the dimension of its structural group is equal to the dimension of $\gal$, i.e., $m(m-1)/2$, ($m=n+1$).
$G$--structures has mathematical interest on its own right (see, for example, 
\cite{Ko}), and we will be interested in two properties of Leibnizian $G$--structures with striking differences in respect to the semi-Riemannian case: their infinitesimal automorphisms (studied below) and the set of all the  compatible affine connections (Section \ref{sZa}). 

An infinitesimal automorphism of a  $G$-estructure is a vector field  $K$  generating a  group of automorphisms of the principal fiber bundle. In the semi-Riemannian case, such a $K$ is called   Killing vector field. In the Leibnizian one, the following definition is equivalent.

\begin{defn} \label{defnkl}
Given $(M,\Omega, \h )$, a vector field $K\in\Gamma (TM)$ is \emph{Leibnizian (Killing)} if its local flows $\psi_s$, preserve the absolute clock and space, that is:
$$
\psi^*_s\Omega=\Omega
 \;\;\;\,\;\;\;and\;\;\;\,\;\;\;
  \psi^*_s \h= \h.
$$
$\lei \equiv {\rm Leib}(M,\Omega , \h )$ will denote the set of all the Leibnizian vector fields.
\end{defn}

As ${\cal L}_K$, the Lie derivative along  $K$, 
 can be recovered from the local flows of $K$, the following 
characterizations of Leibnizian vector fields are straightforward.

\begin{pro} \label{c3.4}
Let $(M,\Omega,\h)$ be a Leibnizian spacetime, and  $K\in\Gamma (TM)$ a vector field. The following assertions are equivalent:

\begin{enumerate}
\item $K$ is a Leibnizian vector field.
\item ${\cal L}_K\Omega=0$ and ${\cal L}_K \h=0$, 

\item The following two properties hold:
\begin{enumerate}

\item $\Omega([K,Y])=K(\Omega(Y))$, $\forall Y\in\Gamma (TM)$ (equally: $-d\Omega(K,Y)= Y(\Omega(K))$).

\item $K\left<V,W\right>=\left<[K,V],W\right>+
\left<V, [K,W]\right>,\;\;\forall V,W\in \Gamma({\rm an}\Omega)$.

\end{enumerate}
\end{enumerate}
In particular, $\lei $ is a Lie algebra.
\end{pro}

\begin{obs} \label{o1} {\rm
(1) The right hand side of  {\it 3(b)} makes sense (i.e.: $[K,V]$, $[K,W] \in \Gamma({\rm an}\Omega)$) when {\it 3(a)} holds.

(2) When $d\Omega = 0$, property {\it 3(a)} holds if and only if  $\Omega(K)=cte$. We will put then, for each $c\in \R $: 
\begin{equation} \label{rl}
{\rm Leib}_c(M) = \{ K \in \lei \mid \Omega (K)=c \}
\end{equation}
(clearly, the relevant cases will be $c=0,1$).

(3) As we will see, the  dimension of $\lei $ may be infinite. 
This was expected from a purely algebraic viewpoint: a straightforward computation  from (\ref{ggal}) shows that the Lie algebra $\gal$ contains elements of rank 1 and, thus, this algebra is of finite type
(see \cite[Proposition 1.4]{Ko}). As a consequence, the automorphisms of a Leibnizian manifold are not necessarily a (finite dimensional) Lie group.
}
\end{obs}

\noindent {\sc B. Fields of Leibnizian Observers.} Consider now the case that $Z$ is a {\it  field of Leibnizian observers (FLO)}, that is, $Z\in \Z $, and $Z$ is Leibnizian\footnote{The name of  {\it rigid vector fields} is also natural for FLO's, see \cite[Section 2.3]{SW}}. We will be interested in the classical interpretations of these vector fields; thus, we assume now $d\Omega =0$. According to formula (\ref{rl}) the set of all de FLO's will be denoted as $\leiu$. 

From Proposition \ref{l2.1}, given $Z\in \Z$ a chart
$(t, x^1,\dots x^n )$  adapted to $Z$ exists. Put: 
$$h_{ij} = \langle \partial_i , \partial_j \rangle, \quad \quad h \equiv \h.$$ 
The following characterization of the FLO's is inmediate from its  definition and Proposition \ref{c3.4}. 

\begin{pro} \label{srl}
Let $(M, \Omega, \h )$ be a Leibnizian spacetime with $d\Omega =0$ and $Z \in \Z$. The field of observers $Z$ is a FLO if and only if  for each $p \in M$ there exists a chart  $(t, x^1,\dots x^n )$ adapted to $Z$ such that:
\begin{equation} \label{hindept}
\partial_t h_{ij} = 0, \quad \forall i,j \in \{ 1,\dots n \} .
\end{equation}
\end{pro}

\begin{obs} \label{obs3.7} {\rm 
Of course, in this case  equality (\ref{hindept}) holds for any chart adapted to $Z$. Thus, the FLO's are those fields of observers satisfying: their observers see that, locally, the metric $\h $ does not change with the local absolute time $t$ (they are always at the same distance of the neighbouring observers).
}
\end{obs}

\noindent {\sc C. Main Result.} Now, let us characterize the dimension of the Lie algebra $\lei $. For simplicity, we will assume the existence of a globally defined time function $T$ (of course, the results hold locally if only  $d \Omega = 0$). 

Notice first that  $\leiu$ may be empty (and then  $\lei = \leic$), no matter the dimension of $\leic$ be. Recall also that a vector field $Z\in \Gamma(TM)$ is called {\it complete} if it admits a globally defined flow $\phi$, i.e., $\phi_t : M \rightarrow M$, for all $t\in \R$ (for $Z \in \Z$, one can say, equally, that the --inextendible-- observers in $Z$ are defined on all  $\R$).

\begin{teo} \label{tlei}
Consider the Leibnizian spacetime $(M, dT, \h)$.

\begin{enumerate}
\item	

\begin{enumerate}
\item Let $K \in \leic$ be. The restriction of $K$ to each hypersurface 
$T \equiv T_0$ (constant) is a Killing vector field of the Riemannian manifold $(T^{-1}(T_0), \h )$. 
\item If $\leic \neq 0$ then {\rm dim}($\leic = \infty$).
\end{enumerate}

\item If $\leiu$ is not empty then it is an affine space of associated vector space $\leic$.

Thus, {\rm dim}$(\lei ) \in \{ 0, 1, \infty \}.$

\item If there exists a complete FLO, $Z \in \leiu$, then:
\begin{enumerate}
\item All the hypersurfaces $T\equiv$ constant are isommetric.
\item If some of them  $T^{-1}(T_0)$ admits a  Killing vector field
$K_0 (\not \equiv 0)$ then dim$\leic = \infty .$
\end{enumerate}
\end{enumerate}
\end{teo}

\dem {\it 1}. Assertion {\it (a)} is obvious. For {\it (b)} take any $K \in \leic$. Notice that, for any function $a: \R \rightarrow \R$ the vector field:
$$
K^a(p) = a(T(p)) K(p), \quad \forall p \in M,
$$
satisfies $K^a \in \leic $ too. If $K \not \equiv 0$, 
one can choose a neighborhood
$U$ where $K$ does not vanish, and some interval   $]T_1 , T_2[, T_1<T_2$ included in $T(U)$. Now, just take infinite independient functions $a(T)$ vanishing outside of $]T_1, T_2[$.

{\it 2}. Obvious.

{\it 3}. For {\it (a)} recall that the flow $\phi_{t}$ of $Z$ generates an isommetry between  $T^{-1}(T_0)$ and $T^{-1}(T_0 + t), \forall t \in \R.$ For {\it (b)}, we have just to find some $K \in \leic$, $K\not\equiv 0$ and apply {\it 1(b)}. Such vector field can be constructed from $K_0$ and the flow of  $Z$ as follows:
\begin{equation} \label{extkil}
K_p  = d\phi_{(T(p)-T_0)}(K_0 [\phi_{-(T(p)-T_0)}(p)])
\end{equation}
(with the notation: $K_0[q]  \equiv (K_0)_q, $ for $q = \phi_{-(T(p)-T_0)}(p)$). $\square $

\begin{obs} \label{obs3.9} {\rm 
Choosing $M= \R \times S$ ($S$ any manifold) with $T: \R \times S \rightarrow \R$ the natural projection, it is not difficult to prove that all the dimensions of Leib$_c(M)$ permited by Theorem  \ref{tlei} can occur. Substracting a small neighborhood of some point, the importance of the hypothesis of completeness in  {\it (3)} can be easily verified, (even though this result is always true locally, for any FLO).

Moreover, locally, when there exist a FLO and there are  $r$ independent Killing vector fields $K_{01}, \dots , K_{0r}$ in the neighbourhood of some point at a hypersurface  $T\equiv T_0$, then infinitely many new FLO's can be constructed,  type $Z^* = Z + \sum_i a^i(T) K_i$, for any functions $a^1, \dots a^r$ and $K_i$'s as in (\ref{extkil}). That is, as the time  $T$ varies, all the observers in $Z^*$ can move in the direction of a spacelike Killing vector field with a speed which depends arbitrarily on $T$; this generalizes well-known properties of the {\em kinematical group}, see \cite{Eh}.


}

\end{obs}


\vspace{3mm}
\begin{quote}
\begin{center}
{\bf {\large
II. GALILEAN STRUCTURES }}
\end{center}
\end{quote}

\section{Galilean spacetimes.} \label{sgal}

{\sc A. Galilean connections.} As already commented, a Leibnizian structure has no canonical affine connection associated. Now, affine connections preserving the Leibnizian structure will be studied. The existence of such a fixed connection can be seen as a physical requirement from gauge covariance. In fact, if no connection is fixed then all the the sections of the principal fiber bundle $GM$, or {\em Galilean reference frames}, are physically equivalent. But, in this case, physical laws as Newton's second one should be covariant under changes of Galilean reference frames. This forces the existence of a gauge field (i.e., a compatible connection) which restates covariance. Recall that General Relativity can also be  seen as a gauge theory, where the gauge invariance under different choices of sections in the principle fiber bundle
of the orthonormal basis, must be preserved. Nevertheless, in this theory the gauge field (the gravitational field) is canonically fixed as the unique torsionless connection of the bundle.
\begin{defn} \label{dcg}
A {\rm  Galilean connection} in a Leibnizian spacetime $(M,\Omega, \h)$, is a connection $\nabla$ such that its parallel transport maps Galilean bases onto Galilean bases. 

A {\rm Galilean spacetime}  $(M, \Omega, \h , \nabla )$ is a Leibnizian spacetime $(M, \Omega, \h)$ endowed with a Galilean connection
$ \nabla $.
\end{defn}
As the connection can be reconstructed from the parallel transport, it is not difficult to check the following characterization.

\begin{pro} \label{c4.11}
An affine connection $\nabla$ on a Leibnizian spacetime  $(M,\Omega,\h)$, is Galilean if and only if  the following two conditions hold:

\begin{enumerate}

\item $\nabla \Omega =0$ (i.e.: $\nabla_X\Omega=0,\;\;\forall X\in\Gamma (TM)$).

\item $\nabla \h =0$, that is: $X\left<V,W\right>=\left<\nabla_XV,W\right>+\left<V,
 \nabla_XW\right>,\;\;\forall X\in\Gamma (TM),\;\forall V,W\in 
\Gamma ({\rm an}\Omega )$.

\end{enumerate}
\end{pro}

\begin{obs} \label{ogal}{\rm
Item {\it 1} holds if and only if 
$\Omega(\nabla_XY)=X(\Omega(Y))$, $\forall Y\in\Gamma (TM)$. Thus, if $\Omega(Y)$ is constant then $\nabla_XY \in \Gamma({\rm an}\Omega), \forall X\in \Gamma(TM)$. In particular, this happens if $Y=Z \in \Z$ or if $Y=V,W \in \Gamma({\rm an}\Omega)$; therefore, the right-hand side of item {\it 2} is well defined.}
\end{obs}
Equally, a Galilean connection can be seen as a connection in the  fiber bundle of the Galilean bases $GM$. As any principal fiber bundle, $GM$ admits connections, but it does not admit necessarily a symmetric connection.
Thus, in principle, Galilean connections are not assumed symmetric. Even more, our results on existence of Galilean connections will be mathematically clearer without this restriction. Thus, the  {\it torsion}
$$
{\rm Tor}(X,Y) = \nabla_X Y - \nabla_Y X - [X, Y],
$$
which measures the lack of symmetry of the connection, will be relevant.
The existence of a symmetric Galilean connection implies restrictions on the  $1$--form $\Omega$, as the following result shows.

\begin{lem} \label{lomegator}
For any Galilean spacetime $(M, \Omega, \h , \nabla )$:
\begin{equation} \label{omegator}
\Omega \circ {\rm Tor} = d\Omega.
\end{equation}
Therefore, if there exist a symmetric Galilean connection then  $d\Omega = 0$.
\end{lem}

\dem By using Remark \ref{ogal}:
\[
d\Omega(X,Y)= 
 X(\Omega(Y))-Y(\Omega(X))-\Omega([X,Y]) 
\]
\[
= 
\Omega (\nabla_X Y) - \Omega (\nabla_Y X) - \Omega ([X,Y])= \Omega 
({\rm Tor}(X,Y))
,\;\,\;\forall X,Y\in\Gamma (TM),  
\]
which proves (\ref{omegator}). $ \square $ 

\begin{obs}{\rm 
If a  $G$--structure is parallelizable then it  admits a symmetric connection  \cite[Proposition 1.2]{Ko}, but the converse is clearly false. Nevertheless, as we will see in  Section \ref{sZa}, if $d \Omega =0$ then there are symmetric connections. Thus, for Leibnizian 
$G$--structures one can say: {\it there exists a symmetric connection if and only if   ``$\Omega$ is parallelizable''} (i.e., locally $\Omega = dt$).
}\end{obs}

\noindent When $d\Omega \neq 0$, only ``connections symmetric for a field of observers'' can be defined: 

\begin{defn} \label{dzs}
Let $Z\in \Z$ be a FO, and  $P^Z $ its associated projection (formula {\rm (\ref{epe})}). A Galilean connection is $Z$--\emph{symmetric}, if:
$$
P^Z \circ {\rm Tor} \equiv 0.
$$
\end{defn}
If $d\Omega = 0$ then $\Omega \circ {\rm Tor} \equiv 0$ and, therefore, 
$P^Z \circ {\rm Tor} \equiv {\rm Tor}$; that is: symmetric and $Z$--symmetric connections are equal. More precisely:

\begin{pro} \label{pzs}
Let  $(M,\Omega, \h, \nabla)$ be a Galilean spacetime. The following assertions are equivalent:
\begin{enumerate}
\item $\nabla$ is symmetric.
\item $d\Omega = 0$ and, fixed any point $p\in M$,  there exist a neighborhood $U$ and a FO on $U$, $Z\in {\cal Z}(U)$ such that $\nabla$ is $Z$--symmetric on $U$.
\item Fixed any point $p \in M$, there exists a neighborhood $U$ and  two FO's $Z, Z'$ on $U$,  which are independent at $p$ and such that $\nabla$ is $Z$ and $Z'$--symmetric on $U$.
\item $\nabla$ is $Z$--symmetric for any FO, $Z\in \Z$.
\end{enumerate}
\end{pro}

\dem By using Lemma \ref{lomegator} and above comments, the implications  $ 1 \Rightarrow 2 \Rightarrow 1 \Rightarrow 4 \Rightarrow 3 $ are obvious. For $3 \Rightarrow 2$, notice that
$$
0 = ( P^Z - P^{Z'}) \circ {\rm Tor}(v,w) = (Z-Z')_p \, d\Omega (v,w),
\quad \forall v, w \in T_pM. \quad \square
$$

\noindent Finally, let us define the following fundamental concepts (see Subsection C for interpretations).

\begin{defn} \label{gravcor}
Let $Z\in \Z$, 
a FO in a Galilean spacetime, $(M,\Omega,\h, \nabla)$. The  \emph{ gravitational field induced by $\nabla$ in $Z$} is the vector field:

\[
{\cal G} = \nabla_ZZ.
\]
The \emph{vorticity} or \emph{ Coriolis field} induced by $\nabla$ in $Z$, is the skew--symmetric two covariant tensor field $\om \equiv \frac{1}{2}{\rm rot} Z$ 
defined by:

\[
\omega(V,W)=\frac{1}{2}\left(\left<\nabla_VZ,W\right>-\left<V,\nabla_WZ \right>\right),\;\,\;\forall V,W\in \Gamma({\rm an}\Omega).
\]
An observer $\gamma: I\rightarrow M$, $\Omega(\gamma') \equiv 1$, is {\emph freely falling} if it is a geodesic for $\nabla$.
\end{defn}

\begin{obs} {\rm 
Recall that  $\Omega ({\cal G})= \Omega (\nabla_Z Z) = Z(\Omega(Z)) = 0$, that is, as the Galilean connection parallelizes  $\Omega$,
the gravitational field is always spacelike.

Analogously, the definition of $\omega$ makes sense because $\omega$ is applied only on spacelike vector fields (Remark \ref{ogal}). In general, the  {\it rotational} of a vector field  rot$X$,  as in  Definition \ref{gravcor}, 
makes sense when  $\Omega(X)$ is constant (in particular, if  $X$ is spacelike or a FO) and it is applied on pairs of spacelike tangent vectors.

} \end{obs}

\noindent {\sc B. Galilean vector fields.} As for the Leibnizian case,   vector fields (and, in particular, FO's) with flows preserving the Galilean structure, becomes natural now. Recall first that, given an affine conection $\nabla$, a vector field $K$ with local flows preserving $\nabla$ (i.e.: ${\cal L}_K \nabla = 0$) is called affine (Killing), and is characterized by the equality:
\begin{equation} \label{afin}
[K, \nabla_Y X] = \nabla_{[K,Y]}X + \nabla_Y [K,X], \quad \forall X, Y \in \Gamma(TM)
\end{equation}
(when $K, X$ and $Y$ are coordinate vector fields, this means that the Christoffel symbols are independent of the coordinate associated to $K$).

\begin{defn}\label{srg}
Given a Galilean structure $\egal $, a vector field
$K \in \Gamma(TM)$ is \emph{ Galilean } (Killing) if $K$ is Leibnizian for
$(M,\Omega, \h)$ and affine for $\nabla$. If, additionally, $K$ is a FO then  $K$ is a  
\emph{field of Galilean observers} (FGO). 
\end{defn}
Denote by Gal$(M) \equiv$ Gal$\egal $ the Lie algebra of all the Galilean vector fields. If $d\Omega = 0$, Gal$_1(M)$ will denote the affine space of all the FGO's, in agreement with the notation in Remark \ref{o1}(2). Although Leibnizian vector fields might have infinite dimension, this cannot hold for the Galilean ones, which are always affine; recall that the maximum dimension for affine vector fields is $m(m+1)$. Therefore, from the classical results by Palais, the diffeomorphisms of $M$ preserving the Galilean structure are a (finite dimensional) Lie group, and its associated algebra is the subalgebra of  Gal$(M)$ generated by its complete vector fields (see, for example, \cite[Vol. I, Note 9]{KoNo}). It is not difficult to find the best bound for the dimension of Gal$(M)$:

\begin{pro} \label{cota}
If $m=$ dim$M$ then dim(Gal$(M)$) $\leq m (m+1)/2$. 
\end{pro}

\dem Choose $p\in M$ and take coordinates $(t, x^1, \dots x^n)$ such that the corresponding set of coordinate vector fields  $(\partial_{\mu})$ is a Galilean base at $p$. Each Galilean vector field $K\in$ Gal$(M)$ is determined by the values of $K^{\mu}(p)$
and $\partial_{\nu} K^{\mu} (p)$\footnote{This holds for  any affine vector field. The proof is analogous to the one for the Killing case in \cite[p. 442-3]{Wald}.}. Condition 
{\it  3(b)} of  Proposition \ref{c3.4} imposes $m (m-1)/2$ independent linear equations for the values of  $\partial_i K^j (p)$; Condition {\it 3(a)}
fixes the values of $\partial_{\nu} K^0, \forall \nu \in \{ 0, 1, \dots , n\}$, that is, it imposes $m$ independent conditions more. $\square$

\begin{obs} \label{ocota} {\rm 
This bound for  dim(Gal$(M)$) is the best one, as one can check in the standard example: $(\R ^{n+1}, dt^2, \h _0, \nabla^0)$, being $t$ the usual projection on the first variable and $\h _0$ (resp. $\nabla^0$) the usual metric on each hypersuperface (resp.  usual connection). 

Remarkably, the maximum dimension of Gal$(M)$ is equal to the maximum dimension for the Killing vector fields of a semi-Riemannian metric on $M$. 
This was expected because, on one hand, the groups $\gal $ and orthogonal $O_s(n+1,\R)$ has the same dimensi\'on and, on the other, Killing vector fields are automatically affine for the Levi-Civita connection of the semi-Riemannian metric.
}\end{obs}
Finally, we give the following consequence on gravitational and Coriolis fields  (Definition \ref{gravcor}), interesting for its classical physical interpretation.
\begin{pro} \label{pincompleta}
Let $Z\in \Z $ be a FGO of $\egal $. Then 
$$
\LL _Z {\cal G} (= [Z, {\cal G}]) = 0, \quad 
\LL _Z \omega = 0, \quad
\LL _Z {\rm Tor}  = 0. 
$$
If $d\Omega =0$, then the first (resp. second, third) equality is equivalent to the following fact: for any chart $(t,x^1,\dots , x^n )$ adapted to $Z$, the field  ${\cal G}$ (resp. 
$\omega$,  Tor) is independent of the coordinate $t$.
\end{pro}

\dem The first equality is a consequence of (\ref{afin}) with $K=X=Y=Z$. From this formula one also has:
\be \label{aux}
[Z, \nabla_X Z] = \nabla_{[Z,X]} Z.
\ee

Then, for any spacelike vector fields $V,W$:
$$ 2 \LL _Z \omega(V,W)= 2\left( Z(\om (V,W)) - \om ([Z,V],W) -\om (V,[Z,W])\right)
$$
$$
= Z\left( \< \nabla_VZ, W\> - \< V, \nabla_WZ\> \right)
$$
$$
-\< \nabla_{[Z,V]}Z, W\> + \< [Z,V], \nabla_W Z\>
-\< \nabla_V Z, [Z,W]\> + \< V, \nabla_{[Z,W]} Z\>.
$$
But this expression vanishes, by using Proposition 3.4 (formula {\em 3(b)}) and (\ref{aux}). For the torsion, we can assume that $X,Y,Z$, at any fixed point, conmute and then:
$$
\LL _Z {\rm Tor}(X,Y) = [Z, \nabla_X Y] - [Z, \nabla_Y X].
$$
By (\ref{afin}), the last two terms vanishes. 

Finally, last assertion is straightforward from the expressions in coordinates. $\square$

\vspace{3mm}

\noindent {\sc C. Classical physical interpretations.} Next, some definitions will suggest the classical interpretations for observers in   $\egal$. For simplicity, we will consider the case $d \Om = 0$ and $\nabla$ symmetric, but the definitions can be extended formally to the general case.
 
Fix a FO,  $Z \in \Z$. Denote, as usual,
$$
A_Z: {\rm an}\Om \rightarrow {\rm an}\Om, \quad A_Z(V) = -\nabla_V Z, \quad \forall V \in \Gamma ({\rm an}\Om ) , 
$$
and decompose $-A_Z $ in its symmetric $\hat S$ and skew-symmetric $\hat \om $ parts\footnote{The sign - in the definition of $A_Z$ is a usual convention  Differential Geometry: $A_Z$ is then the  Weingarten endomorphism for the hypersuperficies  $t\equiv$ constant (see for example \cite{KoNo}). Nevertheless, this sign is ruled out in the decomposition.}.
That is, 
$$
-A_Z = \hat S + \hat \omega 
$$
where $\hat S$ is self-adjoint for $\h$, and $\hat \om$ skew-adjoint. Denote  by  $S, \om$ the corresponding fields of  2-covariant associated tensors:
$$
S(V,W) = \< \hat S(V),W \> =  \frac{1}{2} \left( \< \nabla_V Z, W\> + 
\< V, \nabla_W Z\> \right)
$$
$$
\om (V,W) = \< \hat \om(V),W \> = \frac{1}{2} \left( \< \nabla_V Z, W\> - 
\< V, \nabla_W Z\> \right) .
$$

Tensor $\om$ is, then, the vorticity or Coriolis field in Definition \ref{gravcor}. The name ``vorticity'' means that, if $Z$ represents the trajectories of the particles of a fluid, then $\omega$ measures how, given a fixed trajectory, the others turn around. The name ``Coriolis field'' appears because  $\om $ measures the ``lack of inerciality'' of $Z$ due to the spinning of the observers (even though this lack of inerciality maybe intrinsic, see Remark \ref{o27}). In fact, when $n=3$ and $M$ (or, equally, an$\Om$) is orientable, $\om $ can be represented by a {\it Coriolis vector field} $C_{\om}$ in a standard way. Indeed, fix an orientation continuously at each fiber of an$\Om$; the metric  $\h$ yields a standard {\it oriented volume element}, $dv$, which is a skew-symmetric 3-covariant tensor. Now, define 
$C_{\om}$ by the equality $\om (V,W)= dv(C_{\om}, V, W), \forall V,W \in \Gamma ({\rm an} \Om)$.
$\hat S$ (or,   $S$) will be called the  {\it intrinsic Leibnizian part of $A_Z$}, because of the following result.

\begin{pro} \label{pli}
Fixed $Z \in \Z$, the endomorphism field $\hat S$ (and, thus, $S$) depends only on the Leibnizian structure $(M, \Om , \h )$; thus, it is independent of the Galilean connection $\nabla$.

Moreover, $Z$ is Leibnizian if and only if $\hat S=0$.
\end{pro}
\dem From the definition of $S$ (recall that we assume now Tor$=0$):
\be \label{fpli}
S(V,V)=  \<\nabla_V Z, V\> = \left( \<[V,Z],V\> + \<\nabla_Z V, V\> \right)
= -\<[Z,V], V\> + \frac{1}{2} Z\< V, V\>,  
\ee
and the first assertion holds. Last assertion is straightforward from (\ref{fpli}), the third characterization in Proposition \ref{c3.4}, and Remark \ref{o1}(2). $\square$

\vspace{3mm}

\noindent Now,  $\hat S$ can be decomposed as:
$$
\hat S= \frac{\theta}{n} I + \sigma,
$$
where $I$ is the identity endomorphism, $\sigma$ is the {\it shear}, characterized because it must be traceless, and $\theta $ is the {\it expansion}. So, $\theta$ measures how, fixed an observer, neighboring observers go away on average, and $\sigma$ the deviations of this average.
From  Proposition \ref{pli}, each observer $\gamma$ in a FLO, $Z$, stand at a constant distance from any other observer $\bar \gamma$ in $Z$; nevertheless, depending on the Galilean connection they may rotate  when  $\omega\neq 0$. 
Then, the gravitational field of a FLO $Z$ measures the forces which must be used, in order to compensate gravity and maintain a constant distance between its observers. Alternatively,  $Z$ may represent a  {\it  rigid solid}, and ${\cal G}$ measures  gravitational tensions.

Finally, fields of inertial observers will be defined. Notice that, from a classical physical viewpoint, it is natural to assume that they are FLO's without ``rotations''. But, under our mathematical approach, it is also natural to assume that they are FGO. Thus, we give two definitions.

\begin{defn} \label{sri}
Let $\egal$ be a Leibnizian spacetime with symmetric  $\nabla$, and $Z \in \Z$. We will say that  $Z$ is a {\em   field of inertial observers (FIO)} if $Z$ is a FLO and $\omega =0$.

In this case, the FIO $Z$ is {\em proper} if it is a FGO.
\end{defn}

\section{Existence of Galilean connections:  Fundamental Theorem}
\label{sZa}
   
Next, we determine all  the Galilean connections compatible with a fixed Leibnizian structure. 

Recall that, for a semi-Riemannian metric $g$, all the connections which parallelize $g$ can be computed from their torsion
Tor and Koszul's formula (which determines the Levi-Civita connection, i.e., the unique one with Tor$=0$). The only condition for Tor is to be a 2-skew-symmetric covariant, 1-contravariant tensor field, Tor $\in \Lambda^2(TM, TM)$. Thus, there exists a natural bijection between the connections which parallelize $g$ and the tensors field in $\Lambda^2(TM, TM)$. 

On the contrary,  formula 
(\ref{omegator}) does represent an obstruction for the possible torsions associated to a Galilean connection. As a consequence, we will have to 
consider tensors fields in 
$\Lambda^2(TM,TM)$ under a restriction type (\ref{omegator}). In addition, we will need so many new parameters as restrictions in (\ref{omegator}). As we will see, gravitational and Coriolis fields will be these new parameters.

Our study will be carried out in two steps. In the first one (Subsection A) we will see how, given a Galilean structure and fixed $Z$, the values of
${\cal G}$, $\om $ and  Tor fix the Galilean connection. In the second step (Subsection B) we will see how, given a Leibnizian structure and fixed $Z$, the permitted values of ${\cal G}$, $\om $ and  Tor are in bijective correspondence with the space of all the Galilean connections.
\vspace{3mm}

\noindent {\sc A. Formula ``\`a la Koszul''.} Our aim is to prove formula (\ref{fZa}), which  plays a role similar to Koszul formula in semi-Riemannian Geometry. Our next result is, then, the ``Fundamental Lemma of the Galilean Geometry'' (compare, for example, with \cite[vol. IV, Ch. 6]{Sp}). As previous notation, put, for any Galilean connection $\nabla$,
\be \label{ea}
A(X,Y)= {\rm Tor}(X,Y) + [X,Y] = \nabla_XY-\nabla_YX,\;\,\;\forall X,Y\in\Gamma (TM).
\ee
That is, $A$ is two times the skew-symmetric part of $\nabla$, and it depends just on its torsion. Notice that  $A(Z,W)\in \Gamma({\rm an}\Omega)$, $\forall Z\in \Z, \forall W\in \Gamma({\rm an}\Omega)$ and $A(W_1,W_2)\in \Gamma({\rm an}\Omega)$, $\forall W_1,W_2\in \Gamma({\rm an}\Omega$).

\begin{lem} \label{lfk}
Let $(M,\Omega,\h , \nabla)$ be a Galilean spacetime, and $Z\in \Z$ a FO with gravitational field  ${\cal G}$ and Coriolis $\om$. Then,  $\nabla$ satisfies the following formula:
\[
2\left<P^Z(\nabla_XY),V\right>= X\left<P^Z(Y),V\right>+Y
 \left<P^Z(X),V\right>-V\left<P^Z(X),\,P^Z(Y)\right> 
\]
\[
+ 2 \left( \Omega(X)\,\Omega(Y)\left<{\cal G},V\right>+\Omega(X)\,\omega(P^Z(Y),V)+
 \Omega(Y)\,\omega(P^Z(X),V) \right) 
\]
\[
+ \Omega(X)\left(\left<A(Z,\,P^Z(Y)),V\right> - \left<A(Z,V),\,P^Z(Y)\right>
\right)
\]
\[
-\Omega(Y)\left(\left<A(Z,\,P^Z(X)),V\right>+\left<A(Z,V),\,P^Z(X)\right>
\right)
\]
\begin{equation} \label{fZa}
+\left<A(P^Z(X),\,P^Z(Y)),V\right>-\left<A(P^Z(Y),V),\,P^Z(X)\right>-
 \left<A(P^Z(X),V),\,P^Z(Y)\right>,
\end{equation}
where $X,Y\in\Gamma (TM)$ and $V\in \Gamma ({\rm an}\Om)$ is any spacelike vector field.

\end{lem}

\dem From the cyclic identities:

\be \label{e1}
V\left<P^Z(X),\,P^Z(Y)\right>=
 \left<\nabla_V{P^Z(X)},\,P^Z(Y)\right>+\left<P^Z(X),\nabla_V{P^Z(Y)}\right>
\ee
\be \label{e2}
P^Z(X)\left<P^Z(Y),V\right>=
 \left<\nabla_{P^Z(X)}{P^Z(Y)},V\right>+\left<P^Z(Y),
  \nabla_{P^Z(X)}V\right>
\ee
\be \label{e3}
P^Z(Y)\left<V,\,P^Z(X)\right>=
 \left<\nabla_{P^Z(Y)}V,\,P^Z(X)\right>+\left<V,
  \nabla_{P^Z(Y)}{P^Z(X)}\right>,
\ee
compute (\ref{e2}) + (\ref{e3}) - (\ref{e1}) to obtain:

\[
\left<\nabla_{P^Z(X)}{P^Z(Y)}+
 \nabla_{P^Z(Y)}{P^Z(X)},V\right>=\,P^Z(X)\left<P^Z(Y),V\right>+\,
  P^Z(Y)\left<V,\,P^Z(X)\right>
\]
\be \label{e4}
-V\left<P^Z(X),\,P^Z(Y)\right>- \left<A(P^Z(Y),V),\,P^Z(X)\right>-\left<A(P^Z(X),V),\,P^Z(Y)\right>.
\ee
On the other hand, using (\ref{epe}) and (\ref{ea}):
\[
2\left<\nabla_X{P^Z(Y)},V\right>
= 2\left<\nabla_{P^Z(X)}{P^Z(Y)},V\right> 
+ 2\,\Omega(X)\left<\nabla_Z{P^Z(Y)},V\right>
\]
\be \label{e5}
=\left<\nabla_{P^Z(X)}{P^Z(Y)},V\right> 
+\left<\nabla_{P^Z(Y)}{P^Z(X)},V\right> 
+ \left<A(P^Z(X),P^Z(Y)),V\right>
+ 2\,\Omega(X)\left<\nabla_Z{P^Z(Y)},V\right>.
\ee
Substituing (\ref{e4}) in (\ref{e5}):
\[
2\left<\nabla_X{P^Z(Y)},V\right>
 = \,P^Z(X)\left<P^Z(Y),V\right>+\,
  P^Z(Y)\left<V,\,P^Z(X)\right>
-V\left<P^Z(X),\,P^Z(Y)\right>
\]
\[
- \left<A(P^Z(Y),V),\,P^Z(X)\right>-\left<A(P^Z(X),V),\,P^Z(Y)\right>
+ \left<A(P^Z(X),P^Z(Y)),V\right>
\]
\be \label{e6}
+ 2\,\Omega(X)\left<\nabla_Z{P^Z(Y)},V\right>.
\ee
Substituting also, in the two first terms in the right-hand side of
 (\ref{e6}), the values of $P^Z(X), P^Z(Y)$ by its expresion (\ref{epe}):
\[
2\left<\nabla_X{P^Z(Y)},V\right>=\Omega(X)\left<\nabla_Z{P^Z(Y)},V\right>-\Omega(X)\left<\,P^Z(Y),\nabla_ZV\right>
\]
\be \label{e7}
-\Omega(Y)\left<\nabla_ZV,\,P^Z(X)\right>-\Omega(Y)\left<V,\nabla_{P^Z(X)}Z\right>-\Omega(Y)\left<V,A(Z,\,P^Z(X)\right>+\{ {\rm Koszul} \},
\ee
where:
\[
\{ {\rm Koszul}  \} = 
X\left<P^Z(Y),V\right>+
Y\left<V,\,P^Z(X)\right>-V\left<P^Z(X),\,P^Z(Y)\right>
\]
\[
+\left<A(P^Z(X),P^Z(Y)),V\right>-
\left<A(P^Z(Y),V),\,P^Z(X)\right>-\left<A(P^Z(X),V),\,P^Z(Y)\right>.
\]
But, using $\nabla_X(\Omega(Y)Z) = \Omega(\nabla_XY ) Z + 
\Omega (Y) (\Omega (X) \nabla_Z Z + \nabla_{P^Z(X)}Z)$, one has:
\[
P^Z(\nabla_XY) = \nabla_XY - \Om (\nabla_XY) Z=
\nabla_X(\Omega(Y)Z)+\nabla_X{P^Z(Y)} - \Om (\nabla_XY) Z
\]
\be \label{e8}
= \Omega(X)\Omega(Y){\cal G}+\Omega(Y)\nabla_{P^Z(X)}Z+\nabla_X{P^Z(Y)}.
\ee
Thus, substitute (\ref{e7}) in (\ref{e8}):
\[
2\left<P^Z(\nabla_XY),V\right>=
2\,\Omega(X)\,\Omega(Y)\left<{\cal G},V\right>+\Omega(Y)\left<\nabla_{P^Z(X)}Z,V\right>+\Omega(X)\left<\nabla_{P^Z(Y)}Z,V\right>
\]
\[
+\Omega(X)\left<A(Z,\,P^Z(Y)),V\right>-\Omega(X)\left<\,P^Z(Y),\nabla_VZ\right>-\Omega(X)\left<\,P^Z(Y),A(Z,V)\right>
\]
\[
-\Omega(Y)\left<\nabla_VZ,\,P^Z(X)\right>-\Omega(Y)\left<A(Z,V),\,P^Z(X)\right>-
\Omega(Y)\left<V,A(Z,\,P^Z(X)\right>+
\{ {\rm Koszul} \}
\]
\[
=2\,\Omega(X)\,\Omega(Y)\left<{\cal G},V\right>+2\,\Omega(X)\,\omega(P^Z(Y),V)+2\,\Omega(Y)\,\omega(P^Z(X),V)
\]
\[
+\Omega(X)\left(\left<A(Z,\,P^Z(Y)),V\right>
-\left<A(Z,V),\,P^Z(Y)\right>\right)
\]
\[
-\Omega(Y)\left(\left<A(Z,\,P^Z(X)),V\right>
+\left<A(Z,V),\,P^Z(X)\right>\right)+\{ {\rm Koszul} \},
\]
as required. $\square $

\begin{obs} \label{oza} {\rm
As $\nabla_XY = P^Z(\nabla_XY) + X(\Om (Y)) Z$,  {\em formula {\em (\ref{fZa})} permits to reconstruct  $\nabla$ from $\Om , \h$, {\rm Tor}, and the values of 
${\cal G}, \om$ associated to $Z$}. 
}\end{obs}

\noindent {\sc B.  Natural bijection.} Let us see how, fixed a FO, formula (\ref{fZa}) determines all the Galilean connections of a Leibnizian spacetime. As previous notation, let: 
(i) $\Lambda^2({\rm an}\Omega )$, the vector space of all the 2-covariant skew-symmetric tensors defined on spacelike vectors (that is,  $\vartheta\in\Lambda^2({\rm an}\Omega )$, if and only if , $\vartheta:{\rm an}\Omega\times {\rm an}\Omega \longrightarrow C^\infty(M)$, $\vartheta$ is $C^\infty (M)$--bilinear and skew-symmetric); 
and (ii) $\Lambda^2(TM,{\rm an}\Omega)$, the vector space of all the   2-covariant skew-symmetric tensors, 1-contravariant spacelike valued (that is, $\Theta\in\Lambda^2(TM,{\rm an}\Omega)$, if and only if , $\Theta:\Gamma (TM)\times\Gamma (TM)\longrightarrow \Gamma({\rm an}\Omega)$, $\Theta$ is $C^\infty(M)$--bilinear and skew-symmetric).

\begin{teo} \label{tfun} Given a Leibnizian spacetime $(M,\Omega,\h )$, let ${\cal D}(\Omega,\h )$ be the set of all its Galilean connections. Fixed a FO, $Z$, the map, $D^Z:{\cal D}(\Omega,\h )\longrightarrow \Gamma({\rm an}\Omega)\times\Lambda^2({\rm an}\Omega)\times\Lambda^2(TM,{\rm an}\Omega)$, given by:

\[
D^Z(\nabla)=\left({\cal G} (\equiv \nabla_Z Z),\, \omega (\equiv \frac{1}{2} {\rm rot}Z), \, P^Z\circ {\rm Tor} \right),\;\,\;
  \forall\nabla\in{\cal D}(\Omega,\h ),
\]
is one-to-one and onto.

\end{teo}

\dem Obviously, this map is well-defined. Let us  prove that it is one-to-one. By using (\ref{omegator}), (\ref{ea}), 
\be \label{fa}
P^Z\circ {\rm Tor}=  A(\cdot,\cdot)- d\,\Omega(\cdot,\cdot)Z-[\cdot,\cdot] 
\ee
and: 


\[
D^Z(\tilde\nabla)=\,D^Z(\nabla)\;\,\;\Rightarrow \;\,\tilde{\cal G}={\cal G},
 \,\;\tilde\omega=\omega,
  \,\;\tilde A=A.
\]
Thus, from  formula (\ref{fZa}):

\[
\left<P^Z(\tilde\nabla_XY)-\,P^Z(\nabla_XY),V\right>=0,
 \;\,\;\forall X,Y\in\Gamma (TM),\forall V\in \Gamma({\rm an}\Omega)\;\,\;\Rightarrow 
\]
\[
\tilde\nabla_XY-\nabla_XY=\,P^Z(\tilde\nabla_XY)-\,P^Z(\nabla_XY)=0,
 \;\,\;\forall X,Y\in\Gamma (TM),
\]
as required. 

In order to check that $D^Z$ is onto, fix ${\cal G}\in {\rm an}\Omega$, $\omega\in\Lambda^2({\rm an}\Omega)$ and $\Theta\in\Lambda^2(TM,{\rm an}\Omega)$. Taking into account (\ref{fa}), define:

\[
A(X,Y)=\Theta(X,Y)+d\,\Omega(X,Y)Z+[X,Y], \;\,\;\forall X,Y\in\Gamma (TM).
\]
Then:
\[
\Omega(A(X,Y))= d\,\Omega(X,Y)+\Omega([X,Y])=X(\Omega(Y))-Y(\Omega(X)),
\]
and $A(Z,W)\in \Gamma({\rm an}\Omega)$, $\forall W\in \Gamma({\rm an}\Omega)$, $A(W_1,W_2)\in \Gamma({\rm an}\Omega)$, $\forall W_1,W_2\in \Gamma({\rm an}\Omega)$. As a consequence, there exists an unique map $\Pi:\Gamma (TM)\times\Gamma (TM)\longrightarrow \Gamma({\rm an}\Omega)$, such that:

\[
2\, \left<\Pi(X,Y),V\right>,\;\,\;\forall X,Y\in\Gamma (TM), \forall V\in \Gamma({\rm an}\Omega),
\]
satisfies formula (\ref{fZa}) for previously fixed ${\cal G}$, $\omega$ and $A$. Define then:

\[
\nabla_XY=X(\Omega(Y))Z+\Pi(X,Y), \;\,\;\forall X,Y\in\Gamma (TM).
\]
A straightforward computation shows that the so-defined
$\nabla$ is a  Galilean connection, with $D^Z(\nabla)$ equal to the initial $({\cal G}, \omega, \Theta )$.
 $\square$

 







According to this theorem, there exists a canonical way to construct a Galilean connection from $Z\in \Z$, and a gravitational  and  Coriolis field: the unique $\nabla$ such that
$
D^Z(\nabla)=({\cal G}, \omega, 0).
$
If, additionally, the spacetime satisfies $d\Omega =0$, we can consider only symmetric connections, that is:
\begin{cor} \label{cfunbis}
Let $(M,\Omega,\h )$ be a Leibnizian spacetime, and fix $Z\in \Z$. The set of all the $Z$--symmetric Galilean connections is mapped bijectively onto the set of all the possible gravitational ${\cal G}\in \Gamma({\rm an}\Omega)$ and Coriolis $\om \in \Lambda^2({\rm an}\Omega)$ fields. 

In particular, if $d\Omega=0$ then the set of all the symmetric Galilean connections is also mapped bijectively onto $\Gamma({\rm an}\Omega) \times \Lambda^2({\rm an}\Omega)$.
\end{cor}
Notice also that, when $d\Omega =0$, if non-symmetric connections are considered then Theorem \ref{tfun} can be rewritten putting Tor instead of $P^Z  \circ$Tor.

\begin{obs} \label{ofun} {\rm (1)It is well-known that the set of all the affine connections on a manifold $M$ has a natural structure of affine space, being the associated vector space the one of all the 2-covariant, 1-contravariant tensors fields. As commented at the beginning of this section, if a semi-Riemannian metric $g$ is fixed, the set of all the connections parallelizing $g$ has a natural structure of vector space (the Levi-Civita connection would play the role of vector 0), isomorphic to the vector space of all the possible torsions, i.e., the space $\Lambda^2(TM,TM)$. Recall that $\Lambda^2(TM,TM)$, is a vector fiber bundle, with fiber of dimension $m^2(m-1)/2$. Theorem \ref{tfun} shows that, fixed $Z$, the space
${\cal D}(\Omega,\h )$ admits a natural structure of vector space (the $Z-$symmetric connection with null gravitational and Coriolis fields would play the role of vector 0), isomorphic to the vector space $\Gamma({\rm an}\Omega)\times\Lambda^2({\rm an}\Omega)\times\Lambda^2(TM,{\rm an}\Omega)$. Recall that this vector space is also a vector fiber bundle, with fiber of equal dimension $n + n(n-1)/2 + n^2(n+1)/2 = m^2(m-1)/2$. 

(2) Corollary \ref{cfunbis} can be seen as an improved version of \cite[Theorem 7]{Ku}. In fact, this result asserts that the degrees of freedom for the symmetric Galilean connections can be put in one-to-one correspondence with the set $\Lambda^2(TM)$ of all 2-forms on $M$. Thus, we obtain not only the further splitting of such two forms in ${\cal G}$ and $\omega$ but also the more precise associated physical interpretations, which are developed in the remainder of the  article.}\end{obs}

\section{Formulas for the connection, geodesics and curvature} \label{sfcgc}

Next, we will give explicit formulas in coordinates for the different geometric elements (Christoffel symbols, geodesics, curvature) associated to a Galilean connection. By using Lemma \ref{lfk},  these formulas can be given in terms of the Leibnizian estructure, and the fields ${\cal G}, \omega$, Tor. For simplicity, we will assume that the connection is symmetric and, thus,  $d\Om =0$, but it is not difficult to give general expressions (see the computations below Remark \ref{churri}).

Thus, fix  $\egal$ with a symmetric $\nabla$, and a FO,  $Z\in \Z$. Let
$(t,x^1, \dots , x^n)$ be a chart adapted to 
$Z$ as in Proposition \ref{l2.1}, and let  
${\cal G}^k$ (resp. $\omega_{ij}$)  be the components of the gravitational field  ${\cal G}$ 
(resp.  Coriolis field $\om $) for $Z$. Let $(h^{kl})_{n\times n}$,  be the smooth local functions obtained  from the inverse of the matrix $(h_{ij}=\left<\partial_{i},\partial_{j}\right>)_{n\times n}$ at each point. Indices will be rised as usual, thus 
  $\om_i^k (= \om_i\, ^k) = \sum_l  \om_{il} h^{kl}$

\begin{teo} \label{chris}
The Christoffel symbols of  $\nabla$ in any chart adapted to  $Z\in \Z$ are:

\[
\Gamma_{\mu\nu}^0=0,\;\;\Gamma_{00}^k={\cal G}^k,\;\;
 \Gamma_{i0}^k=
  \omega_i^k+\frac{1}{2}\sum_{l=1}^n h^{kl}\frac{\partial h_{il}}{\partial t},
\]
$\forall\mu,\nu\in\{0,1,\dots , n\},\;\forall i,k\in\{1, \dots , n\}$, 
being the remainder equal to the symbols for the  hypersurfaces  
$t\equiv$constant with the induced metric, i.e.:
\[
\Gamma_{ij}^k=\frac{1}{2}\sum_{l=1}^n h^{kl}\left(
 \frac{\partial h_{il}}{\partial x^j}+\frac{\partial h_{jl}}{
  \partial x^i}-\frac{\partial h_{ij}}{\partial x^l}\right),\;\;\forall i,j,k\in\{1,\dots , n\}.
\]
As a consequence, for any freely falling observer
$\gamma:I\longrightarrow M$ (Definition \ref{gravcor}), 
the following   \emph{equations of the motion} holds, putting $\gamma^i = x^i\circ \gamma$:

\[
\frac{d^2\gamma^k}{dt^2}
+\frac{1}{2}\sum_{i,j,l=1}^n\, (h^{kl}\circ\gamma)
\left(
\frac{\partial h_{il}}{\partial x^j}+\frac{\partial h_{jl}}{
 \partial x^i}-\frac{\partial h_{ij}}{\partial x^l}
\right)\circ\gamma\, \cdot \,
\frac{d\gamma^i}{dt}\,  \frac{d\gamma^j}{dt}
\]

\be \label{geo}
=-{\cal G}^k\circ\gamma -
\sum_{i,l =1}^n (h^{kl}\circ\gamma)\, \left(\frac{\partial h_{il}}{\partial t}
  \circ\gamma\ \right) \, \frac{d\gamma^i}{dt}
-2\sum_{i=1}^n\, (\omega_i^k\circ\gamma)\,
   \frac{d\gamma^i}{dt},
\ee
for all $k\in\{1,\dots , n\}$.
\end{teo}
\dem From Remark \ref{oza}, one has $\Gamma_{\mu \nu}^{0 } = 0$. For the remainder, just apply  formula (\ref{fZa}) with $P^Z(\partial_i) = \partial_i$ and $A(\partial_{\mu} , \partial_{\nu})=0$, (recall that $A(X,Y)=[X,Y], \;\,\;\forall X,Y\in\Gamma (TM),$
because of the symmetry of $\nabla$). $\square$

\vspace{3mm}

\noindent Notice that, if $h_{ij}$ is independent  of $t$ (i.e. $Z$ is a FLO, Proposition \ref{srl}),  the left-hand side of 
(\ref{geo}) yields the acceleration of the curve obtained as the projection of $\gamma$ in an 
 hypersurface $t\equiv$constant (acceleration computed with the metric $\h$ on this  hypersurface). Denote this left-hand side as
$D^h (\gamma^k  )'/dt$.
On the other hand, recall that  $Z$ is an affine vector field if and only if 
$$\partial_t \Gamma_{\mu \nu}^{\rho} = 0,$$ 
for all $\mu, \nu, \rho$. Thus, the following characterization of previously defined field of observers is straightforward (see also  Propositions \ref{srl} and  \ref{pincompleta}).

\begin{cor} \label{ccoor}
Let $\egal$ be a Galilean spacetime with symmetric $\nabla$, and $Z \in \Z$. Then, in the domain of any chart adapted to $Z$:
\begin{enumerate}
\item $Z$ is a FLO if and only if   $\partial_t h_{ij} = 0$. 

In this case, $\Gamma^k_{i0} = \om^k_i $ and, for freely falling observers:

\be \label{geosrl}
\frac{D^h (\gamma^k  )'}{dt}
=-{\cal G}^k\circ\gamma 
-2\sum_{i}^n\,(\omega_{i}^k\circ\gamma)\,
   \frac{d\gamma^i}{dt}.
\ee

\item $Z$ is a FGO if and only if  $\partial_t h_{ij} = \partial_t \om_{ij}= \partial_t {\cal G}^{k} = 0$. 

In this case, {\rm (\ref{geosrl})} holds with ${\cal G}^k = {\cal G}^k (x^1, \dots, x^n), \om^k_i = \om^k_i(x^1, \dots, x^n)$. 

\item $Z$ is a FIO if and only if  $\partial_t h_{ij} = 0, \om_{ij} = 0$. 

In this case, $\Gamma^k_{i0} = 0 $ and, for freely falling observers:

\be \label{geosri}
\frac{D^h (\gamma^k  )'}{dt}
=-{\cal G}^k\circ\gamma.
\ee

\item $Z$ is a proper FIO if and only if  $\partial_t h_{ij} = \partial_t {\cal G}^{k} = 0, \om_{ij} = 0$. 

In this case, {\rm (\ref{geosri})} holds with ${\cal G}^k  = {\cal G}^k (x^1, \dots, x^n)$.
\end{enumerate}
\end{cor}
From the Christoffel symbols one can readily compute the curvature tensor $R$ (we will follow the convention of sign $R(X,Y) = [\nabla_X, \nabla_Y] - \nabla_{[X,Y]}$). As:
\begin{equation} \label{addenda1}
\Omega(R(X,Y)Q)=0, \,\,\,\forall X,Y,Q\in \Gamma(M),
\end{equation} 
the operator $R$ is spacelike--valued; moreover:
\begin{equation} \label{addenda2}
\left<V,R(X,Y)W\right>=-\left<R(X,Y)V,W\right>, \;\;\forall X,Y\in\Gamma (TM), \forall V,W\in {\rm an}\Omega
\end{equation}
(notice that (\ref{addenda1}) and (\ref{addenda2}) are also valid if $\nabla$ is not symmetric). Recall that, in a Galilean spacetime, neither the  4-covariant curvature tensor nor the scalar curvature make sense, but Ricci tensor, Ric, does make sense. For each Riemanian hypersurface $t\equiv$constant, the symbol $\nabla^h$ will denote the Levi-Civita connection (as well as the gradient), and the  corresponding curvature and Ricci tensors (defined on spacelike vectors) will be $R^{h}$, Ric$^h$, resp. If  $R^h \equiv 0$ we will say that the space  $({\rm an}\Om , \h)$ is flat. In this case, if $Z$ is a FLO we can assume that the spacelike coordinates are parallel, i.e., $\Gamma_{ij}^k \equiv 0$ (see  Proposition \ref{eet}  for a  general result).

\begin{cor} \label{ccur}
Given a Galilean spacetime $\egal$ with symmetric $\nabla$, for any chart adapted to $Z \in \Z$ we have:
\begin{enumerate} 
\item $R(\partial_{i} , \partial_{j}) \partial_{k} = R^h(\partial_{i} , \partial_{j}) \partial_{k}$ and {\rm Ric}$(\partial_{i} , \partial_{j})=$
{\rm Ric}$^h (\partial_{i} , \partial_{j})$.  
\item  If $Z$ is a FLO: $R(\partial_{i} , \partial_{t}) \partial_{t} = 
\nabla^h_{\partial_i}{\cal G} - \sum_k \left(\partial_t \om ^k_i + 
\sum_l \om^{l}_i \om^k_l \right) \partial_k.$
(In particular, if $Z$ is a FIO: 
 $R(\partial_{i} , \partial_{t}) \partial_{t} = 
\nabla^h_{\partial_i}{\cal G}$). 

Moreover: {\rm Ric}$(\partial_t , \partial_t) = 
{\rm div}^h {\cal G} + \parallel \om \parallel ^2$, where {\rm div}$^h$ denotes the divergence with respect to $\h$ in the corresponding hypersurface  $t\equiv$constant, and $\parallel \om \parallel ^2 = -\sum_{i,j} \om^i_j \om^j_i$.
(In particular, if $Z$ is a FIO: {\rm Ric}$(\partial_t , \partial_t) = 
{\rm div}^h {\cal G}$).

\item If $Z$ is a FLO:
$R(\partial_{t} , \partial_{i}) \partial_{j} = 
\sum_k \left(-\partial_i \om_j^k   
 +\sum_l (\Gamma_{ij}^l \om_l^k - \Gamma_{il}^k \om_j^l) \right) \partial_k $.

In particular: (a) if $Z$ is a FIO then $R(\partial_{t} , \partial_{i}) \partial_{j} = 0$, and (b) 
if the space is flat, and parallel spacelike coordinates are taken:
$R(\partial_{t} , \partial_{i}) \partial_{j} = 
-\sum_{k} \partial_i \om_j^k   \partial_k. $

\item If  $Z$ is a FLO:
$R(\partial_{i} , \partial_{j}) \partial_{t} = 
\sum_k \left( \partial_i \om^k_j - \partial_j \om ^k_i 
+\sum_l (\Gamma_{il}^k \om_j^l - \Gamma_{jl}^k \om_i^l) 
\right) \partial_k$.

In particular: (a) if $Z$ is a FIO then $R(\partial_{i} , \partial_{j}) \partial_{t} = 0$, and (b) 
if the space is flat, and parallel spacelike coordinates are taken:
$R(\partial_{i} , \partial_{j}) \partial_{t} = 
\sum_k \left( \partial_i \om^k_j - \partial_j \om ^k_i 
\right) \partial_k$.


\end{enumerate}
\end{cor}

\begin{obs} \label{churri} {\rm Item {\it 1} makes natural to define the {\em sectional curvature} of a tangent plane included in an absolute space $\pi_p \subset$ an$\Omega_p$ as the curvature of $\pi_p$ for the  hypersurface $T\equiv T(p)$ endowed with the Riemannian metric $\h$, i.e.
$K(\pi_p) = \langle R^h(v,w)w, v\rangle$, 
where $v,w$ is any orthonormal base of  $\pi_p$. If $\pi_p \subset T_pM$ does not lie in the absolute space an$\Omega_p$, 
we can define:
$$ K(\pi_p) = \langle R(v,Z_p)Z_p, v\rangle,$$
where $v$ is any unit vector of $\pi_p \,  \cap \,$ an$\Omega_p$ and $Z_p\in \pi_p$ satisfies $\Omega(Z_p)=1$.
Thus, from a purely geometrical viewpoint, a rich ``sub-Riemannian'' geometry is introduced in this way,  with interest on its own (compare with \cite{Stri}). 
}\end{obs}

Alternatively, it is not difficult to study the curvature tensor by means of moving frames {\em \`a la Cartan}. For the sake of completeness, we sketch the  structural  equations. Locally, fixed a field of observers $Z$ and an orthonormal base of vector fields
$E_1, \dots, E_n$ of an$\Omega$, consider the  dual base $(\Omega,\varphi^1, \dots,\varphi^n)$ of $(E_0=Z,E_1, \dots, E_n)$, plus the $1$-forms $\varphi^i_{\;\rho}$:

\[
\varphi^i_{\;\rho}(X)=\varphi^i(\nabla_XE_\rho),\;\,\;
 \forall i\in\{1,\dots, n\},\,\forall\rho\in\{0,1,\dots , n\},\,\forall X\in\Gamma (TM).
\]
Then, a straightforward computation shows the following three properties, valid even if $\nabla$ is not symmetric:

\begin{enumerate}

\item The curvature tensor
\[
R(X,Y)E_\rho=\sum^n_{k=1}\Upsilon^k_{\;\;\;\rho}(X,Y)E_k,\;\,\;
\]
is univocally determined by the \emph{Second Structural Equation}:
\[
\Upsilon^k_{\;\;\;\rho}=d\,\varphi^k_{\;\;\;\rho}+\sum^k_{l=1}
 \varphi^k_{\;\;\;l}\wedge\varphi^l_{\;\rho},
  \;\,\;\forall k\in\{1,\dots, n\},\,\forall\rho\in\{0,1,\dots, n\}. 
\]

\item For the gravitational and Coriolis fields ${\cal G}$, $\omega$ of the FO, $Z$, the 
$1$--forms $\varphi^i_{\;0}$  satisfy:
\[
2\;{P^Z}^*\omega={P^Z}^*{\cal G}^\flat\wedge\Omega + 
  \sum^n_{k=1}\varphi^k_{\;\;\;0}\wedge\varphi^k,
\]
where ${\cal G}^\flat(V)=\left<{\cal G},V\right>$, for all $V\in {\rm an}\Omega$.

\item $\Upsilon^j_{\;i}=-\Upsilon^i_{\;j}$ and $\varphi^j_{\;i}=-\varphi^i_{\;j}$, for all $i,j\in\{1,\dots, n\}$. 

Therefore, if $\nabla$ is $Z$--symmetric, the \emph{connection $1$--forms} $\varphi^i_{\;\rho}$, are the unique $1$--forms satisfying the \emph{First Structural Equation}:

\[
\left\{\begin{array}{l}
 2\;{P^Z}^*\omega={P^Z}^*{\cal G}^\flat\wedge\Omega + 
  \sum^n_{k=1}\varphi^k_{\;\;\;0}\wedge\varphi^k \\ [.3cm]
d\,\varphi^i=-\varphi^i_{\;0}\wedge\Omega-\sum^n_{k=1}
  \varphi^i_{\;k}\wedge\varphi^k,                
       \end{array}
\right.
\]
plus the skew--symmetry relations $\varphi^j_{\;i}=-\varphi^i_{\;j}.$
\end{enumerate}


\vspace{3mm}
\begin{quote}
\begin{center}
{\bf {\large
III.  NEWTONIAN STRUCTURES }}
\end{center}
\end{quote}

\section{Newtonian spacetimes} \label{snew}
As a difference with most previous references, our definition of Newtonian spacetime is independent of hypotheses at infinity, i.e., it would be locally testable.
\begin{defn} \label{dnewt}
A Galilean spacetime $\egal$ with symmetric $\nabla$ is \emph{Newtonian} if its space is flat and it admits a FIO.

In this case, the Newtonian spacetime will be  \emph{proper} if some of its FIO's is proper.
\end{defn}

\noindent Now, it is natural to wonder: (A) which hypotheses imply the existence of a FIO? and (B) under these hypotheses, how many FIO's exist? 
In order to answer (A), we will assume for simplicity some global hypotheses, as the existence of a function  absolute time $T$.

\begin{pro} \label{eet}
Let $(M, dT, \h, \nabla) $ be a Galilean spacetime with $\nabla$ symmetric and geodesically complete. Assume that each hypersurface $T\equiv$constant is flat and simply connected. Then:
\begin{enumerate}
\item There exist a FLO, $Z$, and the Leibnizian structure $(M, dT, \h)$ is isomorphic to the standard one
$(\R^{n+1}, dt, \h_0)$ (with $\h_0 = \sum_{i=1}^n (dx^i)^2$ and $(t,x^1,\dots,x^n)$  the usual coordinates of $\R^{n+1}$), being identifiable under the isomorphism $T\equiv t, Z\equiv \partial_t$.
\item Fixed a FLO $Z$ with vorticity $\om$, there exist a FIO (and, then, the spacetime is Newtonian) if and only if there exist a spacelike vector field $A\in \Gamma({\rm an}\Omega)$ such that $2\om = {\rm rot}A$.

Equally, under the identification with $(\R^{n+1}, dt, \h_0)$, there exist a FIO if and only if there exist $n$ functions $a^i: \R^{n+1} \longrightarrow \R$ such that $2\om_{ji} (\equiv 2\om_{j}^i) = \partial_j a^i - \partial_i a^j.$
\item If there exist a FLO, $Z$, with vorticity $\om$ depending only on $T$
(
$\partial_i\om_{jk} \equiv 0$) then there exist a FIO.
\end{enumerate}
\end{pro}

\dem Recall first that  $\Omega(\beta')$ is a constant $c_\beta$ for any geodesic $\beta$. Taking $\beta$ with $c_\beta\neq 0$, the range of $T$ must be all $\R$. By using geodesics with $c_\beta =0$, each hypersurface  $T\equiv$ constant must be isommetric to $\R^n$.

$1.$ The flow $\phi_s$ of $Z$ can be defined directly as follows. Fix any geodesic $\gamma(s)$ parametrized by $T$, i.e.,  $T\circ \gamma (s) = s, \forall s \in \R.$ For each $p\in M$, take the unique spacelike geodesic  $\alpha: [0,1]\rightarrow 
T^{-1}(T(p)) $ connecting $\gamma(T(p))$ with $p$. Let $v_s, s\in \R,$ be the vector field along  $\gamma$ obtained by parallel transport of
$\alpha'(0)$ along $\gamma$, from $\gamma(T(p))$ to $\gamma(T(p)+s)$. If $\alpha^*_s$ is the geodesic with initial velocity $v_s$, define $\phi_s(p) = \alpha^*_s(1)$. 
It is straightforward to check that the infinitesimal generator $Z$ of $\phi_s$ is  a FLO and, fixing an orthonormal base of the absolute space at $\gamma(0)$, the isomorphism with the standard Leibnizian structure is straightforward.

$2.$ Fixed the FLO $Z$ put: $\bar Z = Z - A$, where, using the isomorphism of item {\it 1},
 $A=\sum_k a^k\partial_k$ for some functions $a^k$ on $\R^{n+1}$. Easily, 
 rot$\bar Z(\partial_i, \partial_j) = 
2 \om_{ij} - 
\partial_i a^j  + \partial_j a^i$, and the result follows.

$3.$ Use item {\it 2} with $a^j = -\sum_k \om_{jk} x^k$.
$\square$

\vspace{3mm}

\begin{obs} \label{o27} {\rm 
(1) For all Newtonian spacetimes the Leibnizian structure must be locally isomorphic to the standard one on $\R^m$. For the sake of simplicity,  we will assume from now on that this standard Leibnizian structure underlies globally on any Newtonian spacetime.

(2) From item {\it 2} it is clear that if, for some indexes  $i,j,k$, one has $\partial_i\om_{jk} + \partial_k \om _{ij}
+ \partial_j \om _{ki} \neq 0$ ($\omega$ is not ``spatially closed'') then there are no FIO's.
Notice that, when $Z$ is a FLO but not a FIO: (i) if the spacetime is Newtonian (i.e., there exist a FIO) then $\omega$ represents ``inertial (Coriolis) forces'', (ii) otherwise, $\omega$ represents ``true'' gravitational forces (which cannot be ``gauged away'').

(3) An alternative formulation of Definition \ref{dnewt} is to impose the ``gyroscope principle'': $R(X,$ $Y)V=0$ whenever $V$ is spacelike (see, for example, \cite[Box 12.4, Axiom (3)]{MTW}, \cite[Def. 1.1, Axiom 5]{NS}). In this case,  Corollary \ref{ccur} {\it 1}
implies that the space is flat and Corollary \ref{ccur} {\it 4} plus Proposition \ref{eet} {\it 3} imply the existence of a FIO. 
}\end{obs}

\noindent Next, we will focus on the question (B) at the beginning of this section. Recall first  the following straighforward result.

\begin{lem}\label{lcoi}
Let $(\R^{n+1}, dt, \h_0, \nabla)$ be a Newtonian spacetime and fix a FIO, $Z\in\Z$. Consider a generic FO, 
$\bar Z = Z + \sum_i a^i\partial_i$ for some functions $a^i$ on $\R^{n+1}$.
\begin{enumerate}
\item The relation between the gravitational fields ${\cal G}, \bar {\cal G}$ of $Z$, $\bar Z$ is: 
\be \label{gbar}
\bar {\cal G} = {\cal G} + \sum_{i=1}^n \partial_t a^i \partial_i + \sum_{i,j=1}^n(a^i\partial_i a^j) \partial_j.
\ee

\item $\bar Z$ is a FIO if and only if the $a^i$'s are independent of $x$, $a^i\equiv a^i(t)$ and thus,
\be \label{gbar2}
\bar {\cal G} = {\cal G} + (a^i)'(t) \partial_i.
\ee

\item If $Z$ and $\bar Z$ are proper FIO's then (\ref{gbar}) and (\ref{gbar2}) hold with constant derivatives $(a^i)'$, for all $i$.
\end{enumerate}
\end{lem}
Therefore, if $Z$ is a FIO then $\bar Z= Z+ \sum_i a^i(t) \partial_i$ is a FIO for any $a^i(t)$, and the FIO's have infinite dimension. If $Z$ is proper, $\bar Z$ will be proper 
if and only if $a^i(t) = \alpha^i_1 \cdot t + \alpha_0^i$ for some constants $\alpha_1^i, \alpha_0^i$. And if $Z$ and $\bar Z$ are FIO's (proper or not) with the same gravitational field, then $a^i(t) \equiv \alpha_0^i$ for all $i$. Summing up:

\begin{teo} \label{tcoi}
Let $(\R^{n+1}, dt, \h_0, \nabla)$ be a Newtonian spacetime.
\begin{enumerate}
\item The set of all the FIO's is an affine space of infinite dimension.
\item If the Newtonian spacetime is proper, proper FIO's are a $2n$-dimensional subspace. 
\item Fixed a FIO, $Z$, with gravitational field ${\cal G}$, the set FIO(${\cal G}$)= $\{\bar Z \in \Z | \bar Z \; \hbox{is} $
$ \hbox{a FIO} \hbox{ and} 
\; $
 $\bar {\cal G} 
= {\cal G}\}$ is a $n$-dimensional subspace.

\end{enumerate}
\end{teo}

\begin{obs} \label{pajamental}{\rm 
(1) When $Z$ is a proper FIO, one can also put FIO(${\cal G}$)= $\{\bar Z \in \Z | \bar Z \;$  is a FIO and  $ [Z, \bar Z] = 0\}$. In this case, FIO(${\cal G}$) is the set of all the FO's {\it whose observers move with constant velocity respect to $Z$}. Of course, there are only $n$ independent directions for such  velocities. Any other proper FIO $\bar Z$ measures  a gravitational field $\bar {\cal G} = {\cal G} + {\cal G}_0$, where ${\cal G}_0$ is parallel
(``a uniform gravitational field cannot be distinguised from a uniform acceleration''). 

(2) Any possible gravitational field  ${\cal G}$ for  $\nabla$ fixes the $n$-dimensional set of fields of observers  FIO(${\cal G}$). One of such gravitational fields ${\cal G}_0$ maybe priviledged by some physical or mathematical reason. For example, ${\cal G}_0$ maybe the unique gravitational field vanishing at infinity (this is a natural condition for Poisson's equation) or the unique one vanishing along a concrete observer\footnote{This observer can be called ``the center of the Universe'' following ideas of Newton himself -``the center of the Universe is not accelerated by gravitation''.} $\gamma_0$.
In this case, FIO(${\cal G}_0$) is a distinguished $n$-dimensional set of fields of inertial observers.

(3) It is commonly accepted that ``inertial reference frames'' (see (4) below) can be defined only if there exist a priviledged ${\cal G}_0$ which vanishes at infinity (see for example \cite{Tr}). Under our viewpoint, it is preferable to maintain our definition of FIO's and, when necessary, to speak about proper FIO's or FIO(${\cal G}_0$) (as in the next section). Recall that, under our definition, the question whether a field of observers is inertial or not is {\em purely local} and can be determined, in principle, from Corollaries \ref{ccoor} and  \ref{ccur}. 
At any case, those who prefer more classical names can call our inertial observers ``Newtonian observers'' and reserve the name ``inertial'' for our FIO(${\cal G}_0$) when 
${\cal G}_0$ vanishes at infinity.

(4) From our definition of FIO, we can give a natural definition of {\em inertial reference frame} (IRF), as a particular case of Galilean reference frame (see Section \ref{sgal}A), i.e., as the choice of a priviledged gauge. Consider a Newtonian spacetime, and fix any $p\in M$. Each orthonormal base $(e_1,  \dots , e_n)$  of the absolute space $({\rm an}\Omega_p, \h_p)$ can be parallely propagated to obtain a orthonormal base of vector fields   $(E_1,  \dots , E_n)$. A {\em IRF} is a base of vector fields (moving frame) $(Z, E_1,  \dots , E_n)$ where $Z$ is a FIO and $E_1,  \dots , E_n \in \Gamma({\rm an}\Omega)$ is a parallel orthonormal base of vector fields. The gravitational field of the IRF is, by definition, the one of $Z$\footnote{Notice that this gravitational field is a gauge field; thus, FIO$({\cal G})$ characterizes all the IRF's with the same gauge field ${\cal G}$.}  (the IRF will be {\em proper} if  $Z$ is a proper FIO). Fixed  ${\cal G}_0$, all the IRF's with gravitational field equal to ${\cal G}_0$ are determined by the value of $(Z, E_1,  \dots , E_n)$ at $p$. Thus, the Galilean group $\gal$ acts freely and transitively on the set of all the IRF's with gravitational field ${\cal G}_0$ ({\em classical homogeneous Galilean transformations}).

} \end{obs}

\section{Poisson's equation} \label{spoi}

Up to now, Newtonian spacetimes have been described in a purely geometric way. Notice that the knowledge of a FIO $Z$ and its corresponding ${\cal G}$ allows one to reconstruct $\nabla$ (as a very particular case of formula (\ref{fZa})). Poisson's equation relates  geometry to the ``source'' of the gravitational field, by connecting  ${\cal G}$ to the density of mass. Units with  Gravitational  Newton's constant $G=1$ will be assumed.
 Recall first the following  result (straightforward from  (\ref{gbar2}) and Corollary
\ref{ccur}):

\begin{lem} \label{lpois}
For any Newtonian spacetime:
\begin{enumerate}
\item The spatial divergence of the gravitational field ${\rm div}^h{\cal G}$ is equal for all the FIO's. 

Moreover, Ric$(Z_p,Z_p) = {\rm div}^h{\cal G}(p)$ for all $Z_p$ with $dt(Z_p)=1$ and thus, 
Ric $= 4\pi \rho dt\otimes dt$ where $\rho$ is the {\em density of mass} 
 defined as
$$\rho(t,x) =  {\rm div}^h {\cal G}(t,x)/4\pi .$$ 

\item If, for some FIO $Z$, the gravitational field ${\cal G}$ is a spatial gradient i.e., $
{\cal G}= \nabla^h \Phi $ for some function $\Phi$, 
then the gravitational field $\bar {\cal G}$ of any other FIO $\bar Z = Z + \sum_i a^i(t) \partial_i$ is the spatial gradient
$
\bar {\cal G}= \nabla^h \bar \Phi $ with $$\bar \Phi(t,x)= \Phi(t,x) + \sum_{i=1}^n 
(a^i)'(t) x^i + b^0(t) $$
and $b^0(t)$ arbitrary. 

\end{enumerate}
\end{lem}
Thus, classical Newton's gravitational law and Poisson's equation suggest:
\begin{defn}
A Newtonian (resp. proper Newtonian) spacetime $(\R^{n+1}, dt, \h_0, \nabla)$ is \emph{Poissonian} (resp. {\em proper Poissonian}) 
if the following two conditions hold:

(i) The density of mass is non-negative $\rho \geq 0$.

(ii) The gravitational field
${\cal G}$ of a FIO is a spatial gradient $
{\cal G}= \nabla^h \Phi,
$ for some  $\Phi \in C^{2}(\R^{n+1})$.
\end{defn}

\begin{obs} \label{otr} {\rm
 An alternative assumption to {\it (ii)} is to impose the  conservative character of gravitational forces by means of an assumption on the curvature, say, for some  $Z\in \Z$,
$\langle R(V,Z)Z,W \rangle = \langle R(W,Z)Z,V \rangle $ whenever $V,W$ are spacelike (use Corollary \ref{ccur}; compare with \cite[Def. 1 Axiom 4]{NS},  \cite[Box 12.4, Axiom (7)]{MTW}). From Lemma \ref{lpois}, assumption (i) can also be formulated as Ric$(v,v) \geq 0$ for all $v$. Recall that, at any case,  our axioms avoid any type of redundancy (as, for example, those in \cite[Box 12.4]{MTW}). 
}
\end{obs}
In any Poissonian spacetime,  denoting by $\Delta^h$  the spacelike Laplacian,  intrinsic {\em Poisson's equation}
\be \label{poisson}
\Delta^h \Phi = 4\pi \rho, 
\ee
 hold. Taken coordinates adapted to some FIO $Z$ (and spacelike parallel), it is well--known that if $\Phi(t,x)$ is a solution  of (\ref{poisson}), then $\Phi^*(t,x) = 
\Phi(t,x) + \sum_i b^i(t) x^i + b^0(t) $ is a new solution. Thus, Poisson's equation does not determine univocally the value of ${\cal G}$ for $Z=\partial_t$, but the value of all the possible ${\cal G}$'s for all the FIO's (this happens even in the proper case, where  $\rho$ is necessarily independent of $t$, and the solutions of (\ref{poisson}) can be chosen independent of $t$). But this is not surprising, because, in principle, ({\ref{poisson}) should not priviledge any particular inertial gauge.

In order to avoid this difficulty, one assumes usually that (\ref{poisson}) can be written in coordinates such that $Z=\partial_t$ is not an arbitrary FIO but one in a priviledged 
 set FIO(${\cal G}_0$).
The classical assumption for ${\cal G}_0$ is to assume that it vanishes at spatial infinity (thus, if such a ${\cal G}_0$ exists, then (\ref{gbar2}) implies that it is unique), and this  can  be always assumed if $\rho$ has spatial compact support.

Nevertheless, when $\rho(t,\cdot)$ does not have  compact support for some $t$, 
perhaps no ${\cal G}_0$ vanishes at spatial infinity. The simplest case happens for a
 non-empty spatially homogeneous Universe, i.e., when 
$\rho(t,x) \equiv \rho_0(t)$ with $\rho_0(t) \not\equiv 0$ (even though perhaps 
$\rho_0(t) \equiv $ constant). 
Then, a typical solution of (\ref{poisson}) when $n=3$ is, in spatial spherical coordinates,  $\Phi(t,x) = 2\pi \rho_0(t) r^2 /3$. The corresponding gravitational field ${\cal G}_0$ is null at
$r=0$, i.e., along the observer $\gamma_0(t)= (t,0)$ (the ``center of the Universe'').
Thus, if one choses such a $\gamma_0$, then a tridimensional set of fields of inertial observers 
FIO(${\cal G}_0$) is priviledged, and ${\cal G}_0$ can be reconstructed from $\rho$.

\section*{Acknowledgments}

The authors acknowledge Prof. Mariano Santander from the Universidad de Valladolid for some comments. The second-named author has been partially supported by a  MCyT-FEDER Grant BFM2001-2871-C04-01.


\newpage



\begin{tabular}{|p{1.5in}|p{2.in}|p{2.in}|} \hline

\begin{itemize}
 \item Structure
\end{itemize} & 

 \begin{itemize}{\itemsep 0cm}
  \item[] Semi-Riemannian, $(M,g)$
  \item[] $dim\,M=m$, $index\,g=s$
 \end{itemize} &

 \begin{itemize}{\itemsep 0cm}
  \item[] Leibnizian, $(M,\Omega,<,>)$
  \item[] $dim\,M=m\,(=n+1)$
 \end{itemize}  \\ \hline \hline

\begin{itemize}
 \item Structural group
\end{itemize} & 

 \begin{itemize}
  \item[] Orthonormal, $O_s(m)$
  \item[] $dim\,O_s(m)=m(m-1)/2$
 \end{itemize} & 

 \begin{itemize}
  \item[] Galilean, $\gal$
  \item[] $dim\, \gal=m(m-1)/2$
 \end{itemize} \\ \hline 

\begin{itemize}
 \item Infinitesimal auto\-mor\-phis\-ms
\end{itemize}& 

 \begin{itemize}
  \item[] Killing vector fields
  \item[] Possible dimensions: 
  \item[]  \hspace*{2mm} $0, 1, \dots, \, m(m+1)/2$
 \end{itemize} & 

 \begin{itemize}
  \item[] Leibnizian vector fields
  \item[] Possible dim. ($d\Omega =0$): 
  \item[]  \hspace*{2mm} $0,1$ or $\infty$
 \end{itemize} \\ \hline 

\begin{itemize}
 \item Possible  con\-nec\-ti\-ons $\nabla$ which pa\-ra\-l\-le\-li\-ze the structure
\end{itemize}& 

 \begin{itemize}
  \item[] Determined by all torsi\-on tensors,  
bijective correspondence:  

\hspace*{2mm} Connections $\leftrightarrow$ 
2-covar. 1-contrav. skew-symmetric tensors

  \item[] Unique connection without torsion (Levi-Civita)
 \end{itemize} & 

 \begin{itemize}
  \item[] Determined by: 

\indent (a) Possible torsions: 

\indent \hspace*{2mm} $\Omega\circ$ 
Tor$=d\,\Omega$

\indent (b) Fixed $Z$  ($\Omega(Z)=1$):  

\indent \hspace*{2mm} $\nabla_ZZ$ and $\omega = rot\,Z/2$ 
  \item[] Existence of $\nabla$ without Tor $\Leftrightarrow$ $d\,\Omega=0$
 \end{itemize} \\ \hline 

\begin{itemize}
 \item Fixed a connection $\nabla$ which pa\-ra\-l\-le\-li\-zes the structure
\end{itemize}& 

 \begin{itemize}
  \item[]  Canonically, Tor$=0$
  \item[]  Killing $\Rightarrow$ Affine
  \item[]  No new definition of vector fields required
 \end{itemize} & 

 \begin{itemize}
  \item[] Even if Tor$=0$, 

\hspace*{2mm} Leibnizian $\not\Rightarrow$ Affine
  \item[] Galilean vector fields: 

\hspace*{2mm} Leibnizian $+$ affine
  \item[] Dimension Galilean: 

 \hspace*{2mm} $0, 1, \dots, \, m(m+1)/2$ 
 \end{itemize} \\ \hline 

\end{tabular}

\vspace*{1.2cm}

\begin{center}
Table 1: \\

Semi--Riemannian vs. Leibnizian/Galilean
\end{center}

\end{document}